\documentclass[review, sort&compress]{elsarticle}

\usepackage{lineno, hyperref}
\usepackage{siunitx}
\usepackage{amssymb}
\usepackage{booktabs}
\usepackage{multirow}
\usepackage{subcaption}

\journal{Nuclear Instruments and Methods A}

\begin{document}

\begin{frontmatter}
\title{Rate-capability of the VMM3a front-end in the RD51 Scalable Readout System}

\author[ess,cern,milan]{D. Pfeiffer\corref{cor1}}
\ead{dorothea.pfeiffer@ess.eu}
\author[cern,bonn]{L. Scharenberg\corref{cor1}}
\ead{lucian.scharenberg@cern.ch}
\author[bonn]{P. Schw\"abig\corref{cor1}}
\ead{schwaebig@physik.uni-bonn.de}
\author[ess]{S. Alcock}
\author[cern]{F. Brunbauer}
\author[ess-dmsc]{M. J. Christensen}
\author[bonn]{K. Desch}
\author[cern,hiskp]{K. Fl\"othner}
\author[helsinki]{F. Garcia}
\author[ess,milan]{R. Hall-Wilton}
\author[cern,prague]{M. Hracek}
\author[bnl]{G. Iakovidis}
\author[cern,brussels]{D. Janssens}
\author[bonn]{J. Kaminski}
\author[bonn,hiskp]{M. Lupberger}
\author[cern,bonn]{H. Muller}
\author[cern]{E. Oliveri}
\author[cern]{L. Ropelewski}
\author[srstechnology]{A. Rusu}
\author[ess,cern]{J. Samarati}
\author[cern]{M. van Stenis}
\author[cern]{A. Utrobicic}
\author[cern,uludag]{R. Veenhof}

\cortext[cor1]{Corresponding authors}

\address[ess]{European Spallation Source ERIC (ESS), Box 176, SE-221 00 Lund, Sweden}
\address[cern]{European Organization for Nuclear Research (CERN), 1211 Geneva 23, Switzerland}
\address[milan]{University of Milano-Bicocca, Department of Physics, Piazza della Scienza 3, 20126 Milan, Italy}
\address[bonn]{Physikalisches Institut, University of Bonn, Nu{\ss}allee 12, 53115 Bonn, Germany}
\address[ess-dmsc]{European Spallation Source ERIC (ESS), Data Management and Software Centre, Ole Maal\o{}es Vej 3, 2200 Copenhagen, Denmark}
\address[hiskp]{Helmholtz-Institut für Strahlen- und Kernphysik, University of Bonn, Nu{\ss}allee 14--16, 53115 Bonn, Germany}
\address[helsinki]{Helsinki Institute of Physics, P.O. Box 64, FI-00014 University of Helsinki, Finland}
\address[prague]{Institute of Experimental and Applied Physics, Czech Technical University in Prague,
   Husova 240/5, 110 00 Prague 1, Czech Republic}
\address[bnl]{Brookhaven National Laboratory, PO Box 5000, Upton, NY 11973, United States of America}
\address[brussels]{Physics Department, Vrije Universiteit Brussel, Pleinlaan 2, 1050 Brussels, Belgium}
\address[srstechnology]{SRS Technology, 30 Promenade des Artisans, 1217 Meyrin, Switzerland}
\address[uludag]{Bursa Uluda{\u g} University, G{\"o}r{\"u}kle Kampusu, 16059 Ni{\"u}fer/Bursa, Turkey}

\begin{abstract}
	The VMM3a is an Application Specific Integrated Circuit (ASIC), specifically developed for the
	readout of gaseous detectors.
	Originally developed within the ATLAS New Small Wheel (NSW) upgrade, it has been successfully integrated
	into the Scalable Readout System (SRS) of the RD51 collaboration.
	This allows, to use the VMM3a also in small laboratory set-ups and mid-scale experiments, which make use
	of Micro-Pattern Gaseous Detectors (MPGDs).
	As part of the integration of the VMM3a into the SRS, the readout and data transfer
	scheme was optimised to reach a high rate-capability of the entire readout system and
	profit from the VMM3a's high single-channel rate-capability of $\SI{3.6}{Mhits/s}$.
	The optimisation focused mainly on the handling of the data output stream of the VMM3a, but also
	on the development of a trigger-logic between the front-end cards and the DAQ computer.
	In this article, two firmware implementations of the non-ATLAS continuous readout mode
	are presented, as well as the implementation of the trigger-logic.
	Afterwards, a short overview on X-ray imaging results is presented, to illustrate the high rate-capability
	from an application point-of-view.
\end{abstract}


\end{frontmatter}


\section{Introduction}
A common goal in the development of radiation detectors is to improve their rate-capability.
Prominent examples driving the need for larger statistics, shorter data acquisition times and
better handling of the detector occupancy are the high luminosity upgrades of the
Large Hadron Collider (LHC) in high energy physics or the European Spallation Source (ESS)
for neutron and material sciences.
Alongside the development of new detector technologies coping
with higher event rates comes the development of powerful
readout electronics allowing to deal sufficiently with the
new rate challenges.

This high-rate challenge was one of the reasons for the
development of the VMM3a Application Specific Integrated Circuit (ASIC)
within the ATLAS New Small Wheel Upgrade \cite{vmm3a}.
In parallel to the integration of the VMM3a into the ATLAS
environment, the VMM3a was also integrated \cite{vmm3a-srs}
into the RD51 \cite{rd51} Scalable Readout System (SRS) \cite{srs-ref1},
providing its features for small R\&D setups and
mid-sized experiments.
Since the readout scheme of the VMM3a in the SRS differs from that used by ATLAS, the rate-capability of the VMM3a/SRS system was investigated. The results are presented in this article. In the main part (section \ref{sec:vmm3a-srs}), the readout scheme of the VMM3a within the SRS is described.
The focus is set on a description of the firmware, the
software to control and operate the system, and the data acquisition
rates that can be achieved with well-defined test pulses.
To confirm the test pulse measurements, and to demonstrate the rate performance under application conditions, X-ray measurements were carried out.
The results are described in the second part (section \ref{sec:x-ray-tests}).

\section{The VMM3a ASIC and the RD51 Scalable Readout System}
\label{sec:vmm3a-srs}
\begin{figure}[t!]
    \centering
    \begin{subfigure}{\textwidth}
        \centering
        \includegraphics[width = \columnwidth]{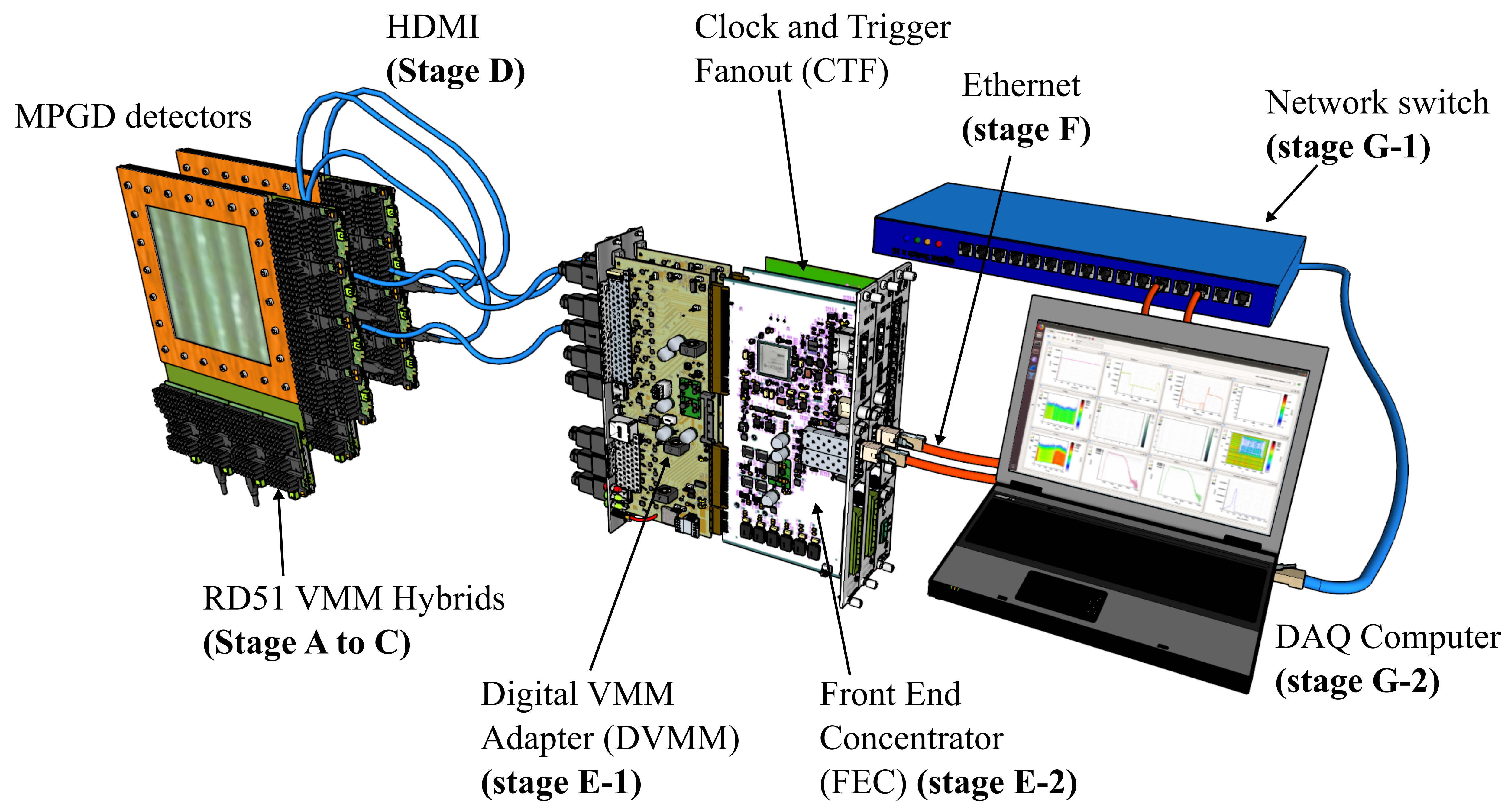}
        \subcaption{3D drawing of the VMM3a/SRS stages.}
        \label{fig:vmm-srs-stages_drawing}
    \end{subfigure}\\
    \vspace{3mm}
    \begin{subfigure}{\textwidth}
        \centering
        \includegraphics[width = \columnwidth]{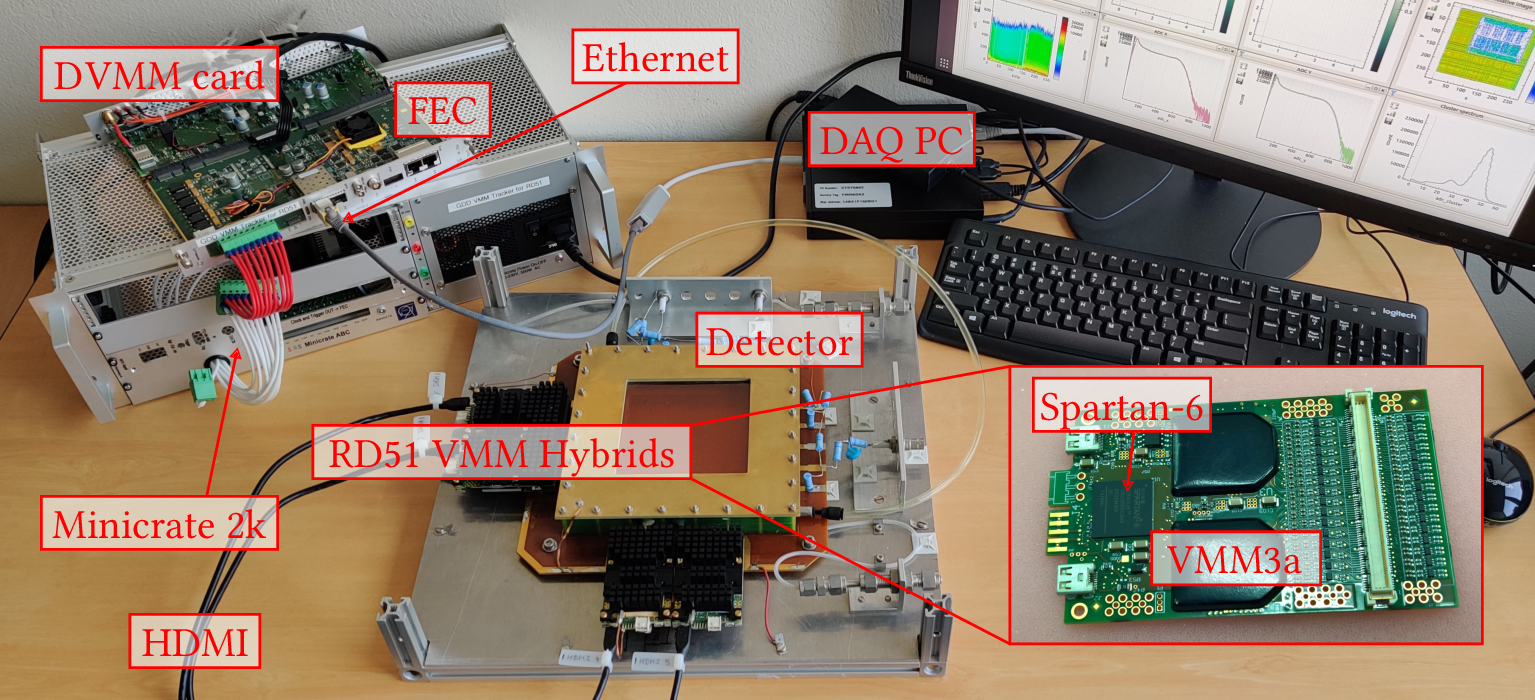}
        \subcaption{Photograph of an R\&D set-up.}
        \label{fig:vmm-srs-stages_photo}
    \end{subfigure}
    \caption{Example of a typical SRS laboratory set-up.
    The RD51 VMM Hybrid contains two VMM3a ASICs (stage A) that
    are connected (stage B) to a Spartan-6 FPGA (stage C).
    Each hybrid is connected via HDMI (stage D) to the combination of DVMM card
    and FEC card (stage E) that here is shown outside the Minicrate 2k.
    The FEC is then connected via Ethernet (stage F) to the DAQ computer (stage G).}
    \label{fig:vmm-srs-stages}
\end{figure}

The Scalable Readout System (SRS) is a versatile and multi-purpose readout system,
developed by the RD51 collaboration for the readout of Micro-Pattern Gaseous Detectors (MPGDs).
A typical configuration of the SRS is shown in figure~\ref{fig:vmm-srs-stages}.
The VMM3a ASIC on the hybrid (stage A) records the charge signals
from the detector readout.
It is connected (stage B) to the Spartan-6 FPGA on the hybrid (stage C) which serially reads the digital
representation of these signals from the VMM3a channels.
It creates hit data from this information, and then sends them out via an HDMI cable (stage D)
to the combination of the Digital VMM adapter (DVMM) card
and the Front End Concentrator (FEC) card (stage E).
The DVMM card has eight HDMI ports, and can also provide power to the hybrids,
either via HDMI cable or via a power cable plugged into its power outlet.
From the FEC card, the data is sent in the form of UDP frames via Ethernet 
(stage F) to a DAQ computer (stage G).

Two different options exist to take data with more than one FEC.
Up to 40 FECs can be connected to the Scalable Readout Unit (SRU),
which has one $\SI{10}{Gbps}$ Ethernet port that communicates with the network card of the
DAQ computer.
Alternatively, the Clock and Trigger Fanout (CTF) card can be used,
which provides a common clock for eight FECs.
For the measurements presented here, a CTF card and two FECs were used.
The data from the two FECs were sent via the FEC $\SI{1}{Gbps}$ network connection
to a $\SI{10}{Gbps}$ switch, and from there to a PC with a $\SI{10}{Gbps}$ network card.

In the following sections, a detailed bottom-up description of the VMM3a ASIC and the SRS is given.
The focus lies on the integration of the VMM3a ASIC into the SRS and the firmware
of the different readout stages.
Further, the software tools needed to control the SRS and to acquire and analyse data are presented.
For a more comprehensive technical description of all components,
please refer to \cite{srs-ref1,srs-ref2} (SRS),  \cite{vmm1,vmm3a} (VMM) and \cite{vmm3a-srs} (VMM3a/SRS).

\subsection{The VMM3a ASIC}
\label{sec:vmm3a-asic}
The VMM3a is a $\num{64}$-channel readout ASIC, developed by the Brookhaven National Laboratory
for the ATLAS New Small Wheel Upgrade, specifically for the readout of gaseous detectors~\cite{vmm3a}.
The analogue part of each VMM channel consists of a charge sensitive amplifier, a shaper,
a discriminator and a peak finder.
The gain of the amplifier can be adjusted within
a range of $\SI{0.5}{mV/fC}$ to $\SI{16.0}{mV/fC}$ (8 discrete settings),
and the peaking time of the shaper can be set to values between $\SI{25}{ns}$ and $\SI{200}{ns}$
(4 discrete settings).
A test pulse generator optionally generates an internal test pulse to inject charge into the
charge sensitive amplifier.
When the signal amplitude crosses the discriminator threshold and the signal peak has been
found by the peak detector (PDO), the charge and time information of the signal is subsequently digitised.
This digitisation or conversion takes $\SI{250}{ns}$.
The maximum rate-capability of a VMM3a channel amounts thus to $\SI{4}{Mhits/s}$.

The peak height of the signal is obtained with a $\SI{10}{bit}$ (effectively $\sim \SI{8}{bit}$)
two-stage current-mode domino ADC~\cite{vmm3a}.
The available time information consists of a $\SI{12}{bit}$ coarse time stamp
and an $\SI{8}{bit}$ fine time stamp. The coarse timestamp, which is called BCID,
is implemented as $\SI{12}{bit}$ Gray counter and counts clock cycles of the
bunch crossing clock CKBC\footnote{The ATLAS experiment at CERN measures particles
	coming from collisions of proton bunches in LHC.
	Every $\SI{25}{ns}$ one proton bunch arrives at the interaction point,
	hence the bunch crossing frequency is $\SI{40}{MHz}$.
	The name bunch crossing clock refers thus to the ATLAS situation,
	but the clock is in reality just a timer/counter and can have any chosen frequency. At a CKBC frequency of $\SI{160}{MHz}$ missing BCIDs were observed. The maximum CKBC frequency that works reliably lies therefore somewhere between $\SI{44}{MHz}$ and $\SI{160}{MHz}$, but has not been determined by the authors. The lowest CKBC frequency is around $\SI{2.5}{MHz}$. When the peak finder finds the peak, the ramp of the TDC starts. The next rising edge of CKBC arms the stop circuit and the next falling edge stops the timing ramp. The TAC slope has thus to be at least $\num{1.5}$ times as long as the period of CKBC. For the longest available TAC slope of $\SI{650}{ns}$, the lowest possible CKBC is thus $\num{1.5}/\SI{650}{ns} = \SI{2.31}{MHz}$. Slower CKBC frequencies therefore mean worse time resolution.}.
The BCID of the digitised signal is the last BC clock before the peak of the signal.
Depending on the chosen BC clock frequency ($\SI{40}{MHz}$ and $\SI[parse-numbers = false]{44.\overline{4}}{MHz}$
in the presented measurements), the resolution of the BCID is $\SI{25}{ns}$ or $\SI{22.5}{ns}$.
The fine time stamp is provided by the time detector (TDO), which uses an
$\SI{8}{bit}$ Time to Amplitude Converter (TAC) with adjustable slopes
($\SI{60}{ns}$ to $\SI{650}{ns}$ in 4 discrete settings).
The measured value represents the time between the peak of the signal and the
falling edge of the next BC clock. Hence in the case of a $\SI{40}{MHz}$ BC clock
and a TAC slope of $\SI{60}{ns}$, the complete time of the peak can be calculated
from the BCID and the TDO with the following formula:
\begin{equation}
\label{eqn:VMM3a_timestamp}
t_{\textrm{chip}} = \textrm{BCID} \times \SI{25}{ns} + (1.5 \times \SI{25}{ns} - \textrm{TDO} \times \frac{\SI{60}{ns}}{255}) \ .
\end{equation} 

\begin{figure}[htb]
    \centering
    \includegraphics[width = 0.9\columnwidth]{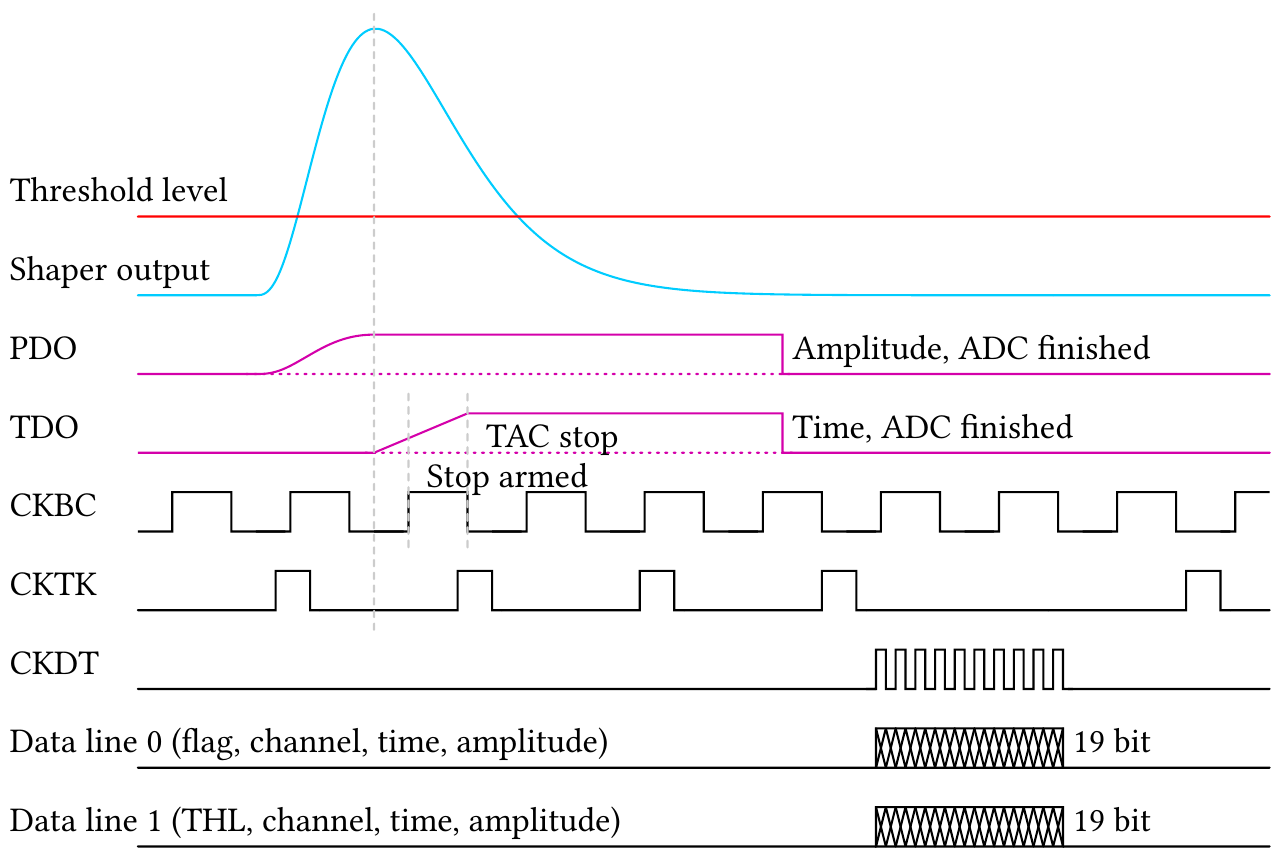}
    \caption{Timing diagram of the VMM3a in the non-ATLAS continuous readout mode.
    	The Peak Detector (PDO) starts the  $\SI{10}{bit}$ ADC for the charge information,
    	and the voltage ramp (TAC slope) for the Time Detector (TDO).
    	This voltage ramp is stopped by the next falling edge of the bunch crossing clock CKBC.
    	The $\SI{8}{bit}$ value of the TDO represents thus the
    	time between the peak and the subsequent BCID.
    	To obtain the digitised data, a token is generated with the token clock CKTK,
    	and passed around to channels with data. The channel that has the token is
    	then bit-wise read out with the data clock CKDT on the two data lines
    	\emph{data\textsubscript{0}} and \emph{data\textsubscript{1}}.}
    \label{fig:timing-diagram}
\end{figure}

The VMM3a configuration data consists of $\SI{1728}{bits}$ and contains global and individual channel settings.
The available per-channel settings are the enabling of the internal test pulse,
the masking of the channel in case of malfunctioning, and finally baseline corrections for the
PDO and the TDO.
Global settings are gain, peaking time, TAC slope and polarity of the input signal.
Further different modes can be chosen, which are not all relevant for the
non-ATLAS continuous mode used by RD51.
Settings that are useful are the neighbouring-logic (also called `neighbour trigger'),
the sub-hysteresis discrimination and the Double Data Rate (DDR) mode for the data clock CKDT.
Channels are normally only read out if the signal
amplitude crossed the discriminator threshold. If the neighbouring-logic (NL)
is enabled, the channels adjacent to channels over threshold are also read out, even when their
signal stayed below the threshold (e.g. channel 8 is over threshold, with the
neighbouring-logic also channels 7 and 9 are read).
The DDR mode for the data clock CKDT means that new bits are available on the data lines \emph{data\textsubscript{0}} and \emph{data\textsubscript{1}} on the rising and the
falling edge of the clock, whereas in single data rate (SDR) mode this happens only on the rising edge of CKDT.
The timing diagram for the data readout of one channel is illustrated in figure \ref{fig:timing-diagram}.

\subsection{The RD51 VMM3a Hybrid}
\label{sec:vmm3a-hybrid}
The RD51 VMM3a hybrid is a small $\SI{5}{cm} \times \SI{8}{cm}$ Printed Circuit Board (PCB) equipped with two VMM3a ASICs, providing thus
128 charge input channels. A 144 pin Hirose connector on the hybrid connects to the readout of a gaseous detector.
From the Hirose connector, the 128 channels are routed to the channel inputs of the two VMM3a ASICs,
with each channel having a spark protection circuit on the PCB.
The other I/Os on the ASICs (clock inputs, configuration inputs, data outputs) are presently connected
to a Xilinx Spartan-6 Field Programmable Gate Array (FPGA).
The FPGA will be upgraded to Spartan-7 in the second half of 2021.
The FPGA provides the necessary clocks for the VMM3a (bunch crossing clock CKBC, token clock CKTK,
data clock CKDT, and the test pulse clock CKTP),
and reads out the two data lines \emph{data\textsubscript{0}} and \emph{data\textsubscript{1}}.
Further, the FPGA takes care of sending the configuration data to the ASICs. 

The format of one $\SI{38}{bit}$ long VMM3a hit (a hit is the data read out from one channel)
is shown in table~\ref{tab:hit}.
\begin{table}[t]
\centering
\caption{Data format of one 38 bit VMM3a hit.}
\label{tab:hit}
\begin{tabular}{l c l}
\toprule
Name & Length (bit) & Values \\
\midrule
\midrule
Data flag & 1 & always 1 \\
\midrule
\multirow{2}{*}{Over-threshold flag} & \multirow{2}{*}{1} & over threshold: 1 \\
& & below threshold (requires NL on): 0\\
\midrule
Channel number &  6 & 0 to 63 \\
\midrule
ADC (PDO) &  10 & 0 to 1023 \\
\midrule
TDC (TDO) &  8 & 0 to 255 \\
\midrule
BCID (clock counter) &  12 & 0 to 4095 \\ 
\bottomrule
\end{tabular}
\end{table}
After reading out the hits from the two VMM3a, they are sent from the hybrid via the DVMM card to the RD51 Front-End Concentrator (FEC) card.
The maximum possible readout rate that can be obtained for the complete VMM3a ASIC depends on the passing speed of the readout token, and the speed of the data transfer. In the following two sections, different firmware approaches to obtain high rates on the hybrid are presented.

The hybrid is connected via HDMI cable to the RD51 FEC.
In the HDMI cable\footnote{The HDMI cable just serves as electrical connection between hybrid and FEC.
The wires of the HDMI cable are not mapped according to the HDMI standard,
and a custom made protocol different from the HDMI protocol is used for data transmission.}
four differential pairs are used for high-speed transmission, and two single wires for I2C.
Via the clock differential pair, a base clock is sent from the FEC to the hybrid.
The purpose of the configuration/trigger pair is to transmit configuration data
and to send control commands like the start and stop of the acquisition from the FEC to the hybrid.
For each of the two VMM3a ASICs on the hybrid, a dedicated differential pair is available for data transmission from hybrid to FEC.
VMM3a data, as well as configuration/trigger data, are $\SI{8}{b}$/$\SI{10}{b}$ encoded.  

All four differential pairs are connected to serialiser/deserialiser (SERDES) components in the FEC and the hybrid firmwares.
In the hybrid firmware, the two data pairs use output serialiser/deserialiser (OSERDES) components,
whereas the clock and configuration/trigger pairs are connected to input serialiser/deserialiser (ISERDES) components.
The ISERDES of the configuration/trigger receiver is used in combination with an input/output delay (IODELAY) component.
The delay of the IODELAY component is automatically adjusted so that each bit is read exactly at the centre of the data eye.
For the correct word alignment of the $\SI{10}{bit}$ word, the bit slip feature of the ISERDES is used.

\begin{figure}[htbp]
    \centering
    \includegraphics[width = 1.0\columnwidth]{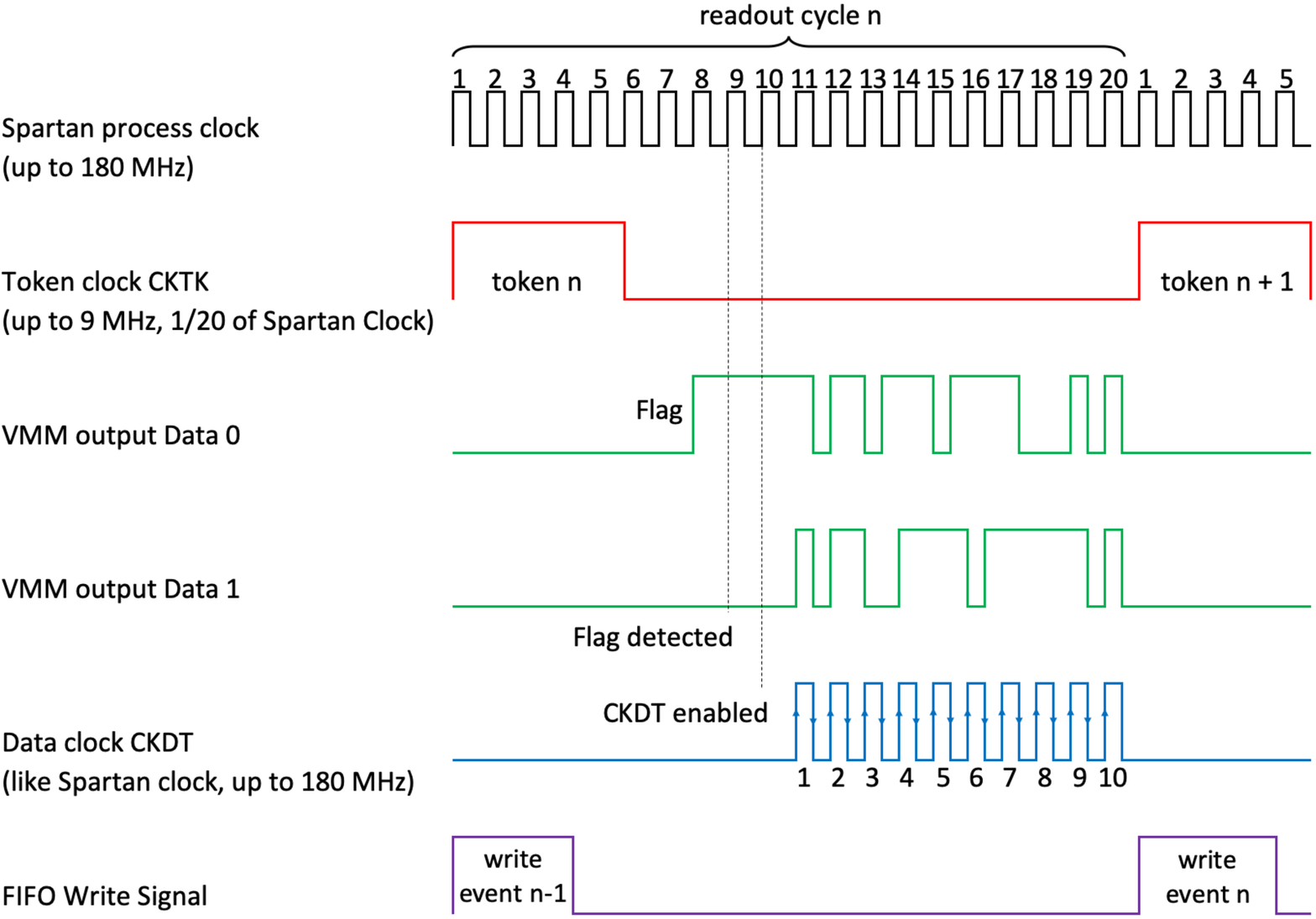}
    \caption{The diagram explains the readout of the VMM3a data on the hybrid FPGA,
    and displays one full readout cycle (cycle \emph{n}).
    In case hit data is available, it is read out in the time between two tokens.
    In the depicted DDR mode, the maximum readout frequency is $\SI{9}{MHz}$,
    whereas in SDR mode it is $\SI{6}{MHz}$.
    Since the time for token generation and readout of the flag remains unchanged,
    the maximum DDR readout frequency is not twice as large as the SDR frequency. }
    \label{fig:ESS_readout-diagram}
\end{figure}

The timing diagram for the data readout from the VMM3a is shown in figure~\ref{fig:ESS_readout-diagram}.
With the help of a counter, the readout token clock CKTK is generated from the readout process
clock with a pulse length of about $\SI{28}{ns}$ and a frequency of up to $\SI{9}{MHz}$.
The token clock is always running and continuously sending tokens.
If data is available, the VMM3a pulls the data line \emph{data\textsubscript{0}} high as a flag.
The firmware detects this flag and starts the data clock CKDT that can have a frequency of
up to $\SI{180}{MHz}$.
Although designed for CKDT frequencies up to $\SI{200}{MHz}$ \cite{vmm3a-srs}, no higher
CKDT frequencies than $\SI{180}{MHz}$ could be reached, due to the observation of occasional data corruption
on some ASICs (see section~\ref{sec:bonn-hybrid-firmware}).

The data clock then pushes out the $\SI{38}{bit}$ VMM3a hit.
Only after the entire hit has been transferred by the data clock, the channel is clear
and can store a new hit\footnote{It should be mentioned that in the non-ATLAS continuous mode,
which is used in the SRS integration of the VMM3a, the 4-hit deep FIFO mentioned in \cite{vmm3a} does not work
(it was identified by the authors together with the designer of the chip \cite{DeGeronimo_email}).
Due to a bug that occurred when the L0 mode was implemented in the 2016 revision of VMM3, the FIFO-read advancement does not work.
This means the next three hits after a hit are stored, but cannot be read out.
The FIFO is not used in the L0 mode used by ATLAS.}.
The diagram depicts the DDR readout mode, with 19 bits coming out on each of the two data
lines \emph{data\textsubscript{0}} and \emph{data\textsubscript{1}}
on the rising and falling edge (when using SDR mode only on the rising edge) of CKDT.
When the complete hit has been read out, the
$\SI{19}{bits}$ from \emph{data\textsubscript{0}} and $\SI{19}{bits}$ from \emph{data\textsubscript{1}} are combined
to a $\SI{38}{bit}$ hit, which is padded with two zeros to $\SI{40}{bits}$.
The $\SI{40}{bit}$ hit is first stored in a $\num{1024}$ hit deep FIFO. This occurs at the same time as the sending of the next token.
The hit data is then read out from the FIFO, encoded in $\SI{8}{bit}$/$\SI{10}{bit}$ format and subsequently sent to the FEC.

\begin{figure}[htbp]
    \centering
    \includegraphics[width = 0.98\columnwidth]{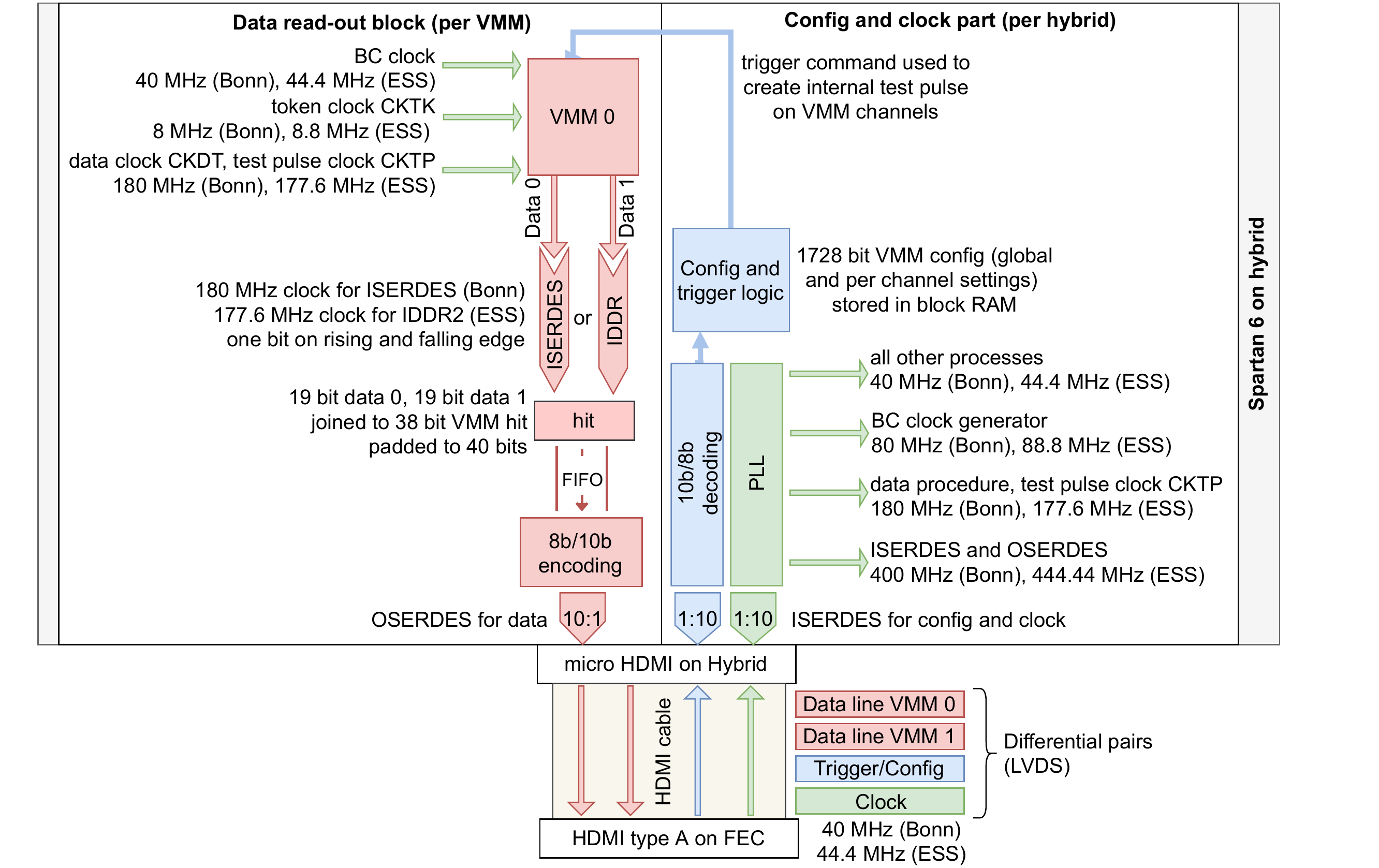}
    \caption{Schematic drawing of the main functional components of the Spartan-6 firmware versions for the RD51 VMM3a hybrid. There are two data readout blocks on the hybrid (one per VMM3a), and a common configuration and clock part.}
    \label{fig:hybrid_logic}
\end{figure}

The following two chapters describe two different high-speed firmware implementations, the Bonn version and the ESS version. The structure of both hybrid firmware versions is shown in figure~\ref{fig:hybrid_logic}. The versions differ mainly in the choice of clock frequencies and the use of different components (ISERDES for Bonn, IDDR for ESS) to read the data from the VMM3a ASIC.

\subsubsection{Bonn firmware for the RD51 VMM3a Hybrid}\label{sec:bonn-hybrid-firmware}
The Bonn firmware of the hybrid was used for initial research of the high-speed readout capabilities of the VMM. It serves as a general-purpose firmware with many configuration possibilities. The base clock of the system is \SI{40}{\mega\hertz}. Using two phase-locked loops (PLLs) all remaining necessary clocks are generated based on the \SI{40}{\mega\hertz} clock. One of the PLLs is used for generating the clocks necessary for data transfers coming from the VMM while the other PLL is used for generating all remaining clocks needed by the firmware. To the latter belongs the bunch crossing clock CKBC which can be chosen from a list of seven different frequencies: \SIlist{2.5;5;10;20;40;80;160}{\mega\hertz}. The incoming data from the VMM3a ASICS is received by ISERDES components with a deserialisation factor of five, reducing the clock frequency behind the IOB by the same factor. This way most of the logic of the firmware uses a lower clock which relaxes timing. To align the data an IODELAY component with fixed tap delay is used in front of the ISERDES.

\begin{figure}[htbp]
    \centering
    \includegraphics[width = 0.9\columnwidth]{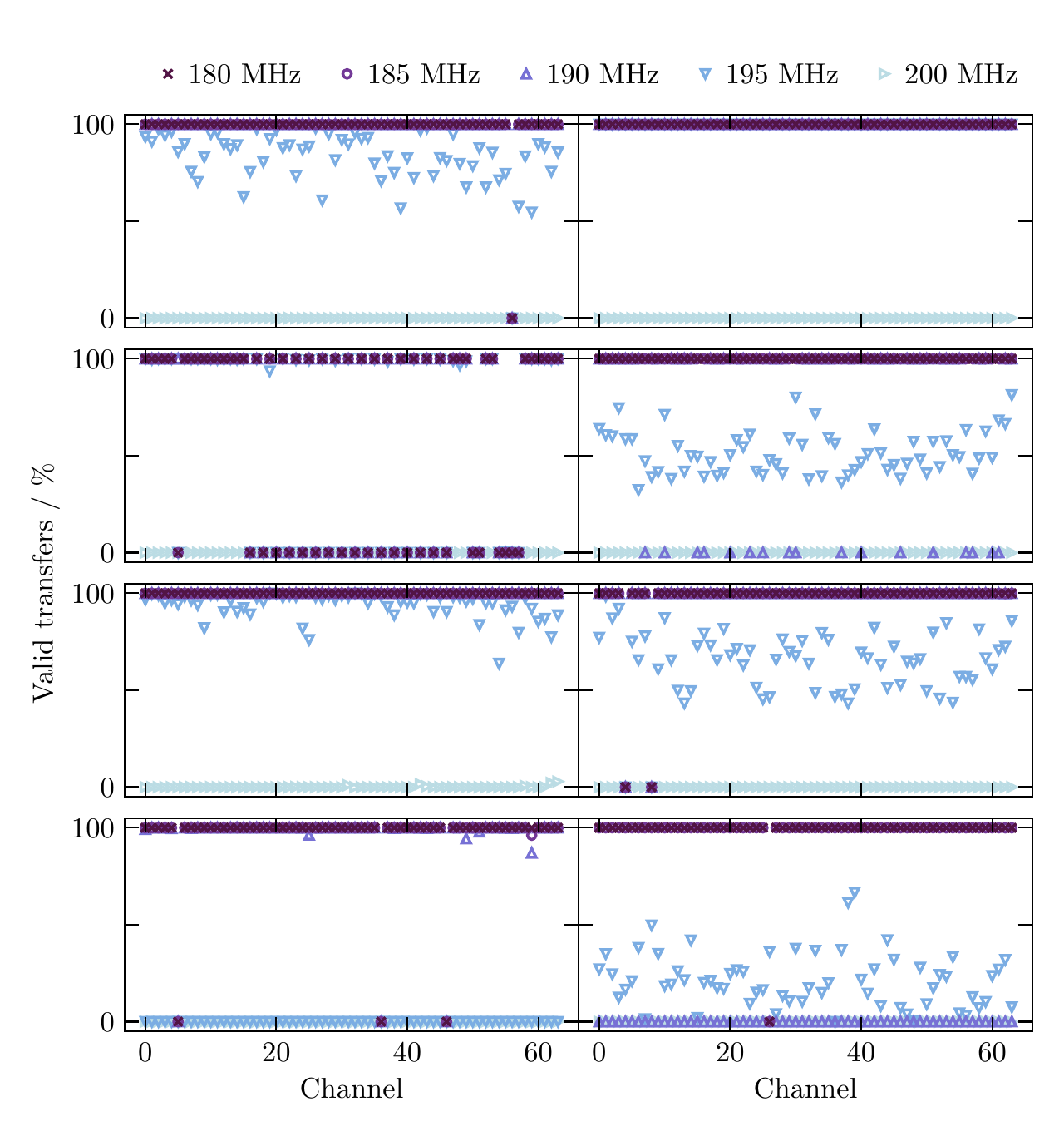}
    \caption{Maximum data transfer clock measurement, results for eight tested VMM3a ASICS and five different transfer clocks CKDT. The relative amount of valid data transfers (with a total of about \num{20000} transfers per VMM3a) is shown for each channel.}
    \label{fig:vmm_rate_compatibility}
\end{figure}

The data transfer clock CKDT has a maximum frequency of \SI{180}{\mega\hertz} which is lower than the design value of \SI{200}{\mega\hertz}, as tests showed that frequencies above \SI{180}{\mega\hertz} resulted in incomplete data transfers when using double data rate. Measurements of this can be seen in figure~\ref{fig:vmm_rate_compatibility}. The lower readout frequency showed to be a design problem of the VMM that was verified in simulation. 
Since all clocks for the data transfers from the VMM3a are generated by the same PLL from which the remaining firmware is independent, it is possible to reprogram the PLL using its Dynamic Reconfiguration Port (DRP) to chose transfer clock frequencies of \SIlist{90;45;22.5}{\mega\hertz} instead.

When using data transfers with double data rate the token clock CKTK
has a frequency of \SI{8}{\mega\hertz} with a duty cycle of roughly
\SI{20}{\percent} and is generated using a counter.
The resulting token length of roughly \SI{28}{\nano\second}
was tested to be necessary such that all tested VMM3a ASICs recognise the token.

The number of active channels has a significant influence on the highest possible data rate. This is due to an empty data transfer which occurs when the token has to be passed to a channel with lower channel number: when there is a data transfer from channel $n$ but no channel with number $> n$ has data to transfer, the token must be passed down to a lower channel or has to stay at $n$. The latter is the case if no other channel has any data to be transferred. The VMM3a will then require an additional token on CKTK during which it does not shift out any data, effectively reducing the data rate of the VMM. Having $N$ active channels and a sufficiently high hit rate the total transfer time for all channels is given by $(N+1)\times t_\textrm{d}$ with $t_\mathrm{d} = 1/f_\textrm{CKTK}$. In order to improve the data rate, the firmware will increase the token clock CKTK and send the next token after a shorter time $t_\textrm{p}$ if the VMM3a does not initiate a data transfer some time after the previous token. The highest expected hit rate considering the effects of the additional token and influence of active channel count and varying token clock is given by
\begin{equation}\label{eq:max_hit_rate}
    \textrm{hit rate} = \frac{\textrm{active channels}}{\textrm{active channels}\times t_\textrm{d}+t_\textrm{p}}\times\frac{\textrm{acquisition window}}{4095} \ ,
\end{equation}
with the acquisition window being a configurable amount of counts by the Gray counter for which the VMM is enabled during a complete BCID cycle.

\begin{figure}[htbp]
    \centering
    \includegraphics[width = 0.9\columnwidth]{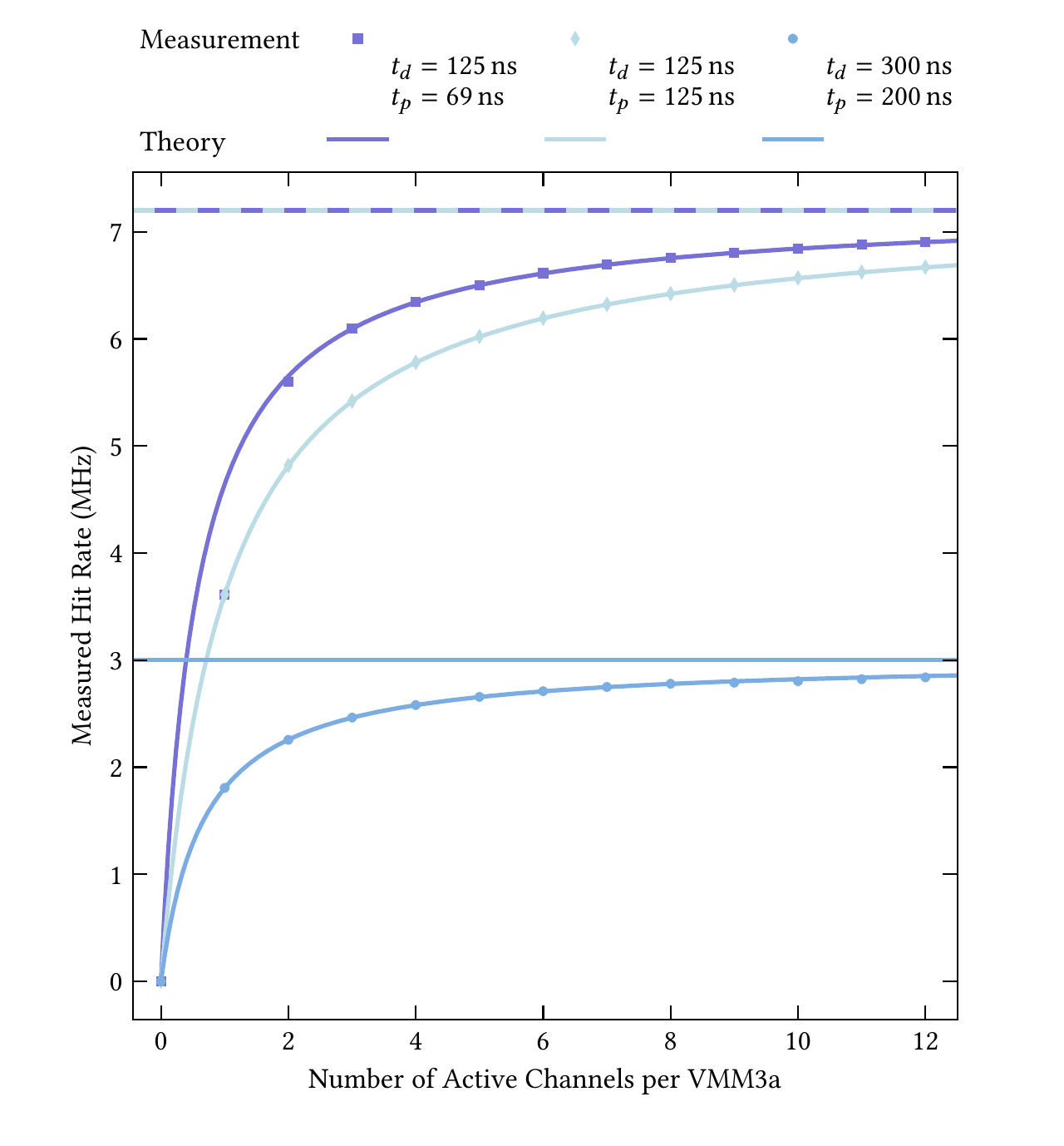}
    \caption{The hit rate received from a single VMM as a function of the number of active channels for three different Bonn firmware versions (similar limitations apply to the ESS firmware, but are not shown here). The markers show the measurement and the lines the expected rate calculated based on the times $t_d$ and $t_p$. Note that for a single active channel the measurements for the upper two curves are identical. The horizontal lines mark the rates in case the hit rate would be independent of the amount of active channels. Stimulus for all measurements is an externally generated, periodic sawtooth waveform.}
    \label{fig:rate_channel_dependence}
\end{figure}

Measured hit rates in dependence on the number of active channels as well as the corresponding expected rate calculated using equation~\ref{eq:max_hit_rate} are shown in figure~\ref{fig:rate_channel_dependence}. It can be seen that theoretical limit and measurement are in very good accordance, with the extrapolated maximum hit rate close to $\SI{7.2}{\mega hits/s}$. When looking at the rates for one active channel note that the upper limit for a single channel is \SI{4}{Mhits/s} although in tests no VMM performed with rates above \SI{3.6}{Mhits/s}. Since this upper limit varied between different VMM ASICs it is assumed that this is due to the ASIC itself. Tests with different waveforms and added time offsets between channels did not increase the measured hit rate.

An additional feature of the firmware is the auto alignment: for updates and changes of the firmware -- especially those regarding the readout and clocking scheme -- the data read from the VMM3a might become misaligned such that bits are missing at the beginning or end. Therefore the firmware offers the possibility to automatically align the \SI{38}{bit} data word correctly.

\subsubsection{ESS firmware for the RD51 VMM3a Hybrid}
\label{sec:ess-hybrid-firmware}
The implementation of the data readout in the ESS firmware is depicted in figure~\ref{fig:hybrid_logic}.
Via the clock differential pair, a $\SI[parse-numbers = false]{44.\overline{4}}{MHz}$
base clock is sent from the FEC to the hybrid. The $\SI[parse-numbers = false]{44.\overline{4}}{MHz}$ clock from the FEC is fed into a phase-locked loop,
which creates all the clocks needed in the firmware\footnote{When the hybrid is used as part of the
ESS readout and not the SRS, the basic clock frequency send to the hybrid will be $\SI{44.0265}{MHz}$,
half of the ESS facility clock frequency of $\SI{88.053}{MHz}$.}. 
The output clocks are distributed via global clock buffers (BUFG) components.
The general firmware logic uses a frequency of $\SI[parse-numbers = false]{44.\overline{4}}{MHz}$.
The clock generator for the bunch crossing clock works with $\SI[parse-numbers = false]{88.\overline{8}}{MHz}$
and generates a fixed $\SI[parse-numbers = false]{44.\overline{4}}{MHz}$ CKBC.
The data readout procedure and the test pulse clock generator all use $\SI[parse-numbers = false]{177.\overline{6}}{MHz}$.
The fastest clock is the $\SI[parse-numbers = false]{444.\overline{4}}{MHz}$ clock for the SERDES components.  

The clock and configuration part of the firmware is common for both ASICs,
whereas one data readout block has been implemented per VMM3a.
The two data lines (\emph{data\textsubscript{0}} and \emph{data\textsubscript{1}}) of the VMM3a are connected
to input double data rate register (IDDR2) components.

Assuming the clock settings described above (the $\SI[parse-numbers = false]{44.\overline{4}}{MHz}$ clock from the FEC
is taken as basic clock for the hybrid), the maximum hit rate is $\SI[parse-numbers = false]{8.\overline{8}}{Mhits/s}$.
This rate is perfectly matched to the speed of the SERDES.
With a clock frequency of $\SI[parse-numbers = false]{444.\overline{4}}{MHz}$ and $\SI{8}{b}$/$\SI{10}{b}$ encoding,
the effective bit rate of the SERDES amounts to 
\begin{equation}
\label{eqn:serdes_bit_rate}
\textrm{bit rate}_{\textrm{max}} = \SI[parse-numbers = false]{444.\overline{4}}{Mbps} \times 0.8 = \SI[parse-numbers = false]{355.\overline{5}}{Mbps} \ .
\end{equation} 
For a $\SI{40}{bit}$ hit, the hit rate is hence 
\begin{equation}
\label{eqn:vmm_hit_rate}
\textrm{hit rate}_{\textrm{max}} = \SI[parse-numbers = false]{355.\overline{5}}{MHz}/\SI{40}{bits} = \SI[parse-numbers = false]{8.\overline{8}}{Mhits/s} \ .
\end{equation}
This is close to the maximum of $\SI{9}{Mhits/s}$, which can be read from a single VMM3a.
This maximum rate would require a SERDES clock of $\SI{450}{MHz}$ and a CKDT of $\SI{180}{MHz}$.

\subsection{The RD51 SRS FEC}
\label{sec:srs-fec}
\begin{figure}[htbp]
    \centering
    \includegraphics[width = 0.9\columnwidth]{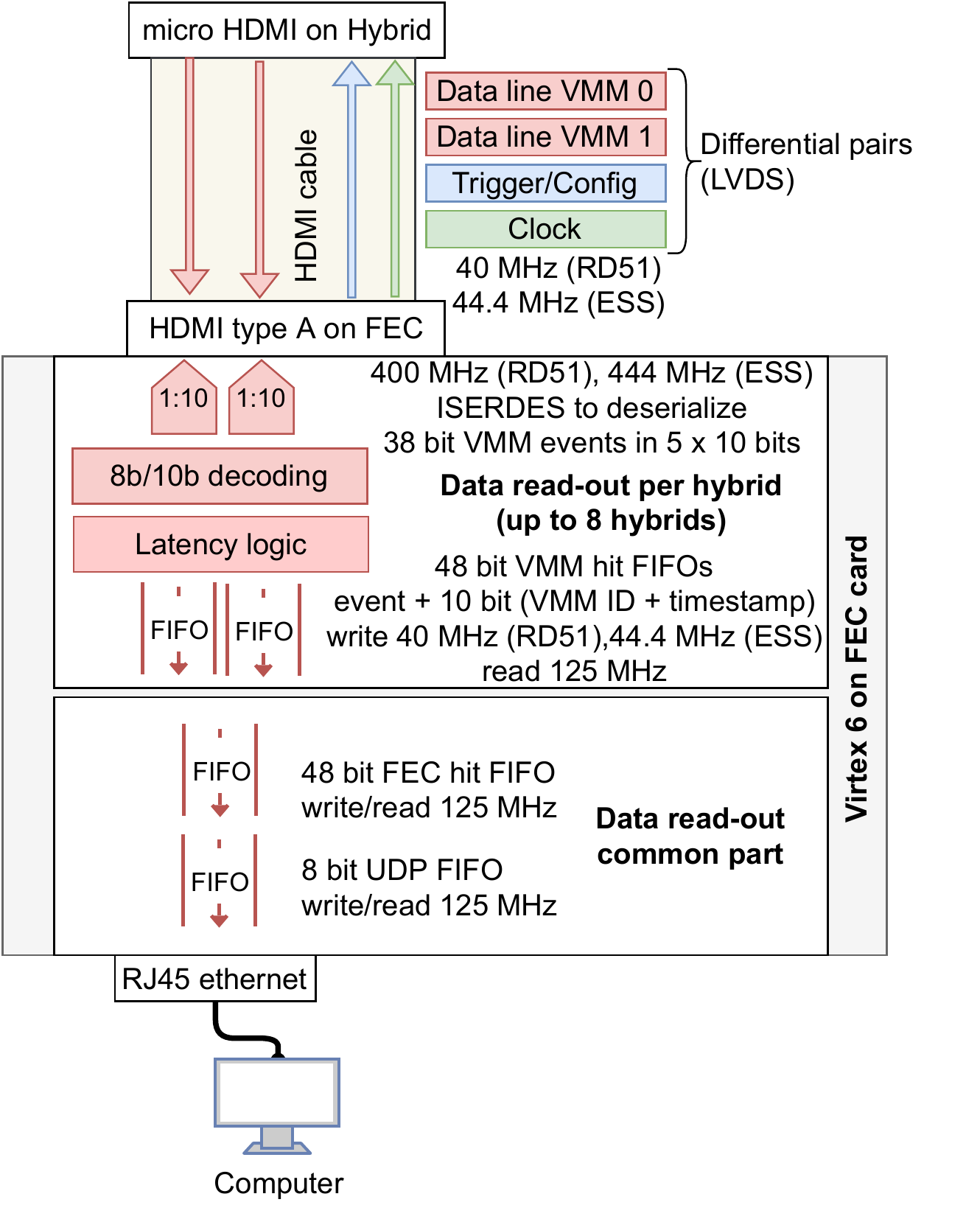}
    \caption{Schematic drawing of the data path in the Virtex 6 firmware of the RD51 SRS FEC.}
    \label{fig:ESS_fec_logic}
\end{figure}

The combination of RD51 FEC and DVMM card can read out up to 8 RD51 VMM3a hybrids.
The hybrids are connected via HDMI cables to the DVMM card,
from which the signals are routed to a Xilinx Virtex 6 FPGA on the FEC.
Figure~\ref{fig:ESS_fec_logic} explains the data path of the FEC firmware.

The differential pairs in the HDMI are described in chapter~\ref{sec:ess-hybrid-firmware}.
The ESS version of the FEC firmware operates the four SERDES in DDR mode with a $\SI[parse-numbers = false]{222.\overline{2}}{MHz}$ clock,
and sends a $\SI[parse-numbers = false]{44.\overline{4}}{MHz}$ clock to the hybrids, whereas the Bonn version uses $\SI{200}{MHz}$
and sends a $\SI{40}{MHz}$ clock. The bit rate of each SERDES is thus $\SI[parse-numbers = false]{444.\overline{4}}{Mbps}$ for the ESS version, and $\SI{400}{Mbps}$ for the Bonn version. Due to $\SI{8}{b}$/$\SI{10}{b}$ decoding, the effective incoming bit rate on each data line on the FEC is reduced by \SI{20}{\percent} to $\SI[parse-numbers = false]{355.\overline{5}}{Mbps}$~\ref{eqn:serdes_bit_rate}
and $\SI{320}{Mbps}$, respectively.
After $\SI{8}{b}$/$\SI{10}{b}$ decoding, the hits from the VMM3a are then time-stamped
on the FEC in the latency logic block, and subsequently as $\SI{48}{bit}$ hits written into a hit FIFO.
For each hybrid connected to the FEC, the firmware provides two separate FIFOs (one per VMM3a).
Whenever a hit is available, it is written with a $\SI[parse-numbers = false]{44.\overline{4}}{MHz}$ write clock to the FIFO. 

In the readout part of the firmware, there is one common FEC hit FIFO for all hits.
A fair scheduler runs with $\SI{125}{MHz}$ and reads one hit at a time from all the non-empty VMM3a data FIFOs.
Provided that it is not full, the hits are transferred to the FEC FIFO.
To be sent out via UDP, the $\SI{48}{bit}$ hits have to be decomposed into six $\SI{8}{bit}$ chunks
and written to the UDP FIFO. The 8 bits from the UDP FIFO are read with $\SI{125}{MHz}$,
as determined by the $\SI{1}{Gbps}$ Ethernet of the FEC.
Since the maximum data rate per VMM3a is around $\SI{356}{Mbps}$,
with the theoretical maximum at $\SI{360}{Mbps}$, the FEC can only support
two VMM3a or one RD51 hybrid at maximum rates.

\subsubsection{Latency logic on the FEC}
\label{sec:fec_latency_logic}
The coarse time stamp in the hit data, the BCID, is a 12-bit counter and overflows every 4096 BC
clock periods or $\SI{92.16}{\micro s}$~\footnote{To simplify the description,
in the following the $\SI{22.5}{ns}$ BC clock period of the ESS version is taken for the calculations.
For the Bonn version, one simply has to replace the $\SI{22.5}{ns}$ with $\SI{25}{ns}$.}.
The FEC generates a $\SI{42}{bit}$ FEC time stamp with $\SI{22.5}{ns}$ resolution,
which only overflows every $\SI{1.2}{days}$. To save bandwidth, the $\SI{42}{bit}$ timestamp
is only send every 16 BCID overflow periods of $\SI{92.16}{\micro s}$, i.e. every $\SI{1.47}{ms}$ for each VMM.
This combination of $\SI{42}{bit}$ timestamp, $\SI{5}{bit}$ VMM-ID and a $\SI{1}{bit}$ flag (always 0)
is called \emph{marker}.

In the time between markers, the $\SI{5}{bit}$ VMM-ID and a $\SI{5}{bit}$ overflow cycle or offset counter
are added to the $\SI{38}{bit}$ VMM data.
The DAQ can distinguish between markers and data by looking at the $\SI{1}{bit}$ flag.
If the flag is zero it is a marker, if it is one it is a hit.

The hits from the VMM3a are read out and transmitted in a roughly time ordered way.
The readout token inside the VMM3a is passed around between channels containing data
in a round-robin like fashion, going from lower channel numbers to higher channel numbers.
Therefore slight disturbances of the time order can occur, as illustrated with the following example:
the ASIC registered  three hits that occurred within one BC clock cycle.
Channel $\num{2}$ has a hit with BCID $\num{0}$, whereas channel $\num{1}$ and channel $\num{63}$ have a hit with BCID $\num{4095}$.
After reading channel $\num{1}$ (BCID $\num{4095}$), the token will be passed to channel $\num{2}$ (BCID $\num{0}$),
and subsequently to channel $\num{63}$ (BCID $\num{4095}$). For the hit in channel $\num{2}$ with BCID $\num{0}$,
the BCID overflow already happened.
The BCID in channel $\num{1}$ and $\num{63}$ has not overflown yet, since the hit occurred slightly earlier.
When the FEC generates the higher-order time stamp, the firmware has thus to consider that
the hit with BCID $\num{0}$ did not occur $\num{4094}$ clock cycles earlier than the hits with BCID $\num{4095}$,
but instead slightly later. 

To be able to make the decision to which overflow cycle a hit belongs,
the FEC also counts BC clock cycles, similar to the BCID counter in the VMM3a.
After receiving the acquisition start signal from slow control, the FEC starts incrementing
its BC counter and then sends via the hybrid a soft reset command to the VMM3a.
This soft reset command sets the BCID counter on the VMM3a to zero.
After sending the soft reset, the FEC is ready to accept data from the hybrids.
The time at which the FEC sends the soft reset is called \emph{reset latency} and can be adjusted via the slow control.

\begin{figure}[htbp]
    \centering
    \includegraphics[width = 1.0\columnwidth]{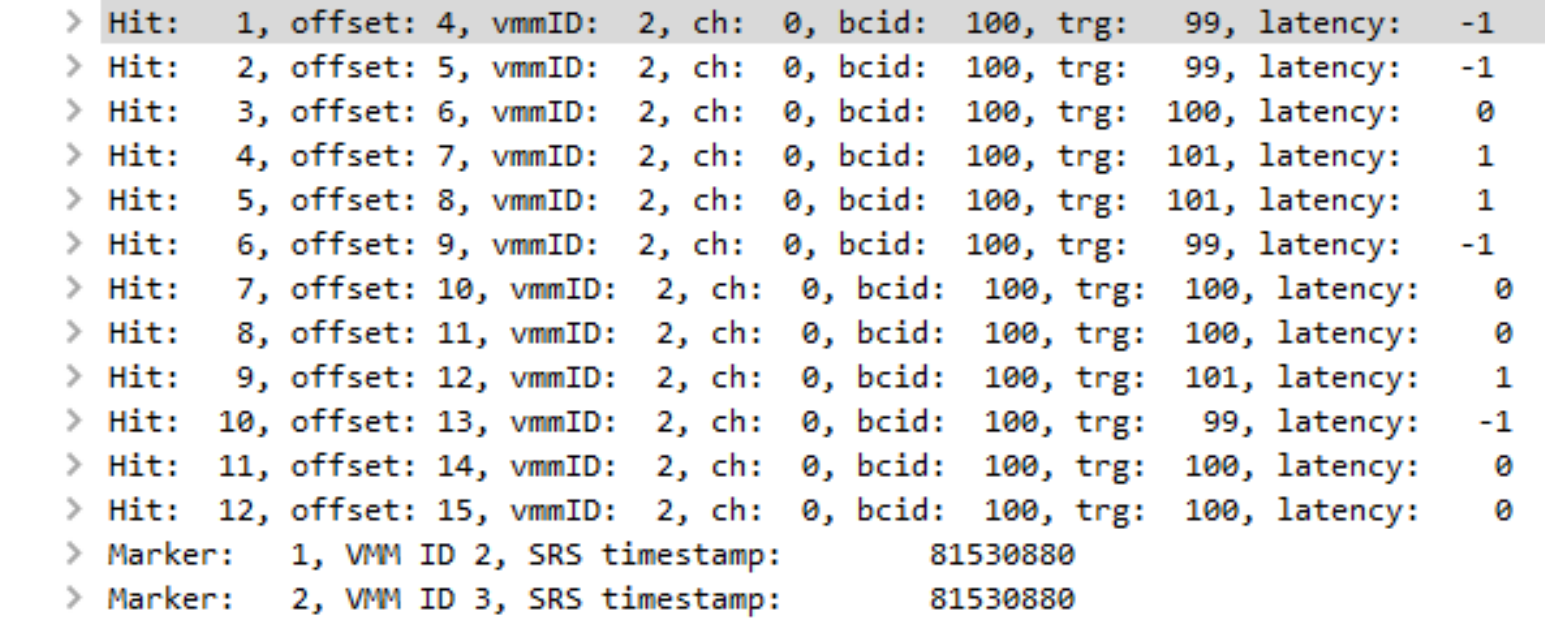}
    \caption{Channel 0 on VMM 2 is pulsed once every 4096 BC clock cycles at BC clock 100. The Wireshark trace shows the hit data with the offset period and the BCID. The field \emph{trg} represents the value of the FEC BC counter. The difference between the \emph{trg} and the BCID is the latency, which has values here between -1 and 1. After all data from offset 15 is transmitted, a marker with the 42-bit SRS time is sent for every active VMM.}
    \label{fig:reset_latency}
\end{figure}

This \emph{reset latency} time is comprised of two components:
the time that it takes the soft reset command to reach the VMM3a and reset the BCID,
and the time that it takes for a single hit from the VMM3a to reach the FEC.
If correctly set, a single hit with e.g. BCID $\num{100}$ will arrive on the FEC at the moment where the FEC BC counter is also $\num{100}$. The aim is thus to have the same values in BCID and FEC BC. With $\SI{2}{m}$ HDMI cables, the \emph{reset latency} latency is of about $\num{47}$ clock cycles. That means the soft reset is sent when the FEC BC counter has a value of $\num{4096} - \num{47} = \num{4049}$. After setting the reset latency to the correct value,
the FEC BC counter is now identical to the BCID of a single hit upon arrival on the FEC. Figure \ref{fig:reset_latency} shows data from one channel on one VMM acquired with the correct latency settings.

A small jitter of a few clock cycles can nevertheless occur. This parameter is called \emph{latency jitter} and is usually set to around $\num{4}$ clock cycles.
The third important parameter is the \emph{maximum latency}.
If all $\num{64}$ channels are activated at the same time,
there is a substantial difference in data arrival time on the FEC between the first
and the last channel that is read out.
On the hybrid one hit can be read out every $\num{5}$ clock cycles of the BC clock or every $\SI{112.5}{ns}$.
So when reading out $\num{64}$ hits that have been digitised at the same time,
the last hit will arrive with a maximum latency of $\SI{7.2}{\micro s}$.

\begin{figure}[htbp]
    \centering
    \includegraphics[width = 1.0\columnwidth]{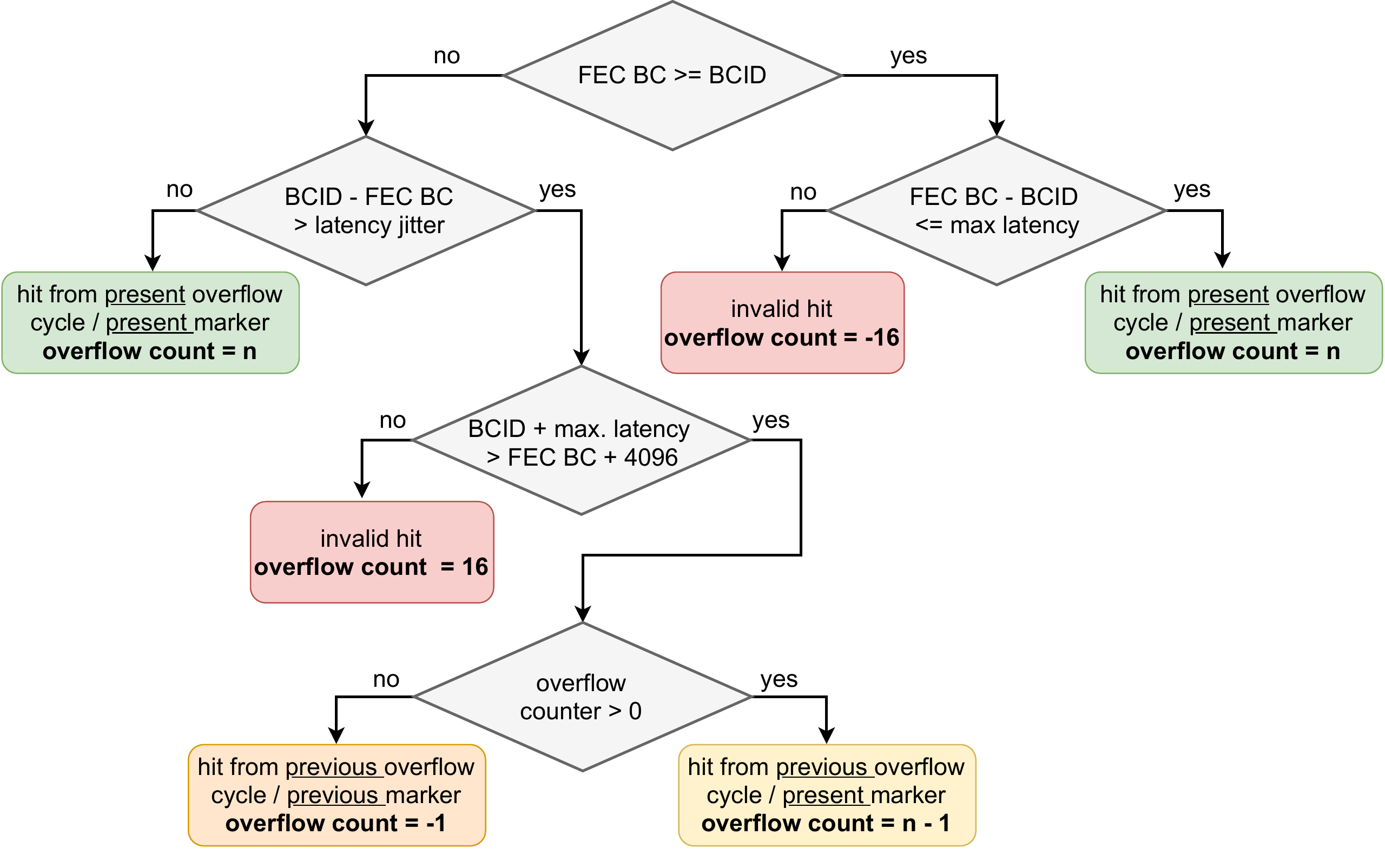}
    \caption{The latency logic flow chart explains how the FEC firmware determines
    the correct higher order time stamp for the hits that arrive on the FEC.
    The BCID of the hit is compared with the FEC BC counter.
    Depending on the parameters latency jitter and maximum latency,
    an overflow counter is added to the hits.
    In the most common case the hit belongs to the present overflow cycle (cases in green boxes).
    If the FEC BC is a lot smaller than the BCID (FEC BC counter plus $\num{4096}$ is smaller
    than the BCID plus the maximum latency), then the FEC BC counter has already overflown
    while data of the previous overflow cycle are still read out.
    The previous overflow cycle can either belong to the present marker (yellow box),
    or be the last overflow cycle of the previous marker (orange box).
    In case the latency conditions are violated, the hit is marked as invalid.
    With correctly set parameters and working hardware this case does not occur.}
    \label{fig:FEC_latency}
\end{figure}

Figure~\ref{fig:FEC_latency} explains how the latency logic is implemented.
In addition to counting BC cycles, the FEC also counts the overflows of its BC counter.
After $\num{16}$ overflow cycles, a new marker is generated and written into each VMM FIFO.
The overflow counter is then reset from $\num{15}$ to zero.
Upon arrival on the FEC, the BCID of a hit is compared with the FEC BC counter.
Based on the result of the comparison,
the FEC firmware can decide to which overflow cycle a hit belongs.
A hit can belong either to the present overflow cycle, the previous overflow cycle
or violate the latency conditions.
In the easiest and most common case, the hit belongs to the present overflow cycle.
The present overflow counter is then added to the data.
If the hit belongs to the previous overflow cycle, the overflow counter is decremented by one.
Therefore overflow counters with values between $-\num{1}$ and $+\num{15}$ can be added to the valid data.
Invalid hits that violate the latency conditions are marked with an overflow counter of $-\num{16}$.
With a correctly configured system and fully operational hardware, invalid hits do not occur. 

The latency logic makes it possible to continuously take data without applying
an acceptance window and without rejecting data.
If measuring e.g. X-rays or white noise, the time distribution of the acquired data is flat.
In the DAQ or the data analysis, the correct FEC timestamp is always calculated from
the last marker that has been sent.
The overflow periods are added to the marker (or subtracted from the marker
if the overflow count is $-\num{1}$) according to the formula:
\begin{equation}
\label{eqn:FEC_timestamp}
t_\textrm{FEC} = t_\textrm{marker} \times \SI{22.5}{ns} + \textrm{overflow counter} \times 4096 \times \SI{22.5}{ns} \ . 
\end{equation} 
To obtain the complete hit time stamp, the FEC timestamp (equation~\ref{eqn:FEC_timestamp})
and the VMM3a timestamp (equation~\ref{eqn:VMM3a_timestamp}) have to be added together:
\begin{equation}
\label{eqn:hit_timestamp}
t_\textrm{hit} = t_\textrm{FEC} + t_\textrm{VMM3a} \ .
\end{equation} 

\subsection{The DAQ PC}
\label{sec:daq-pc}
The RD51 slow control tool \emph{vmmdcs} configures FEC, hybrid and VMM3a~\cite{vmmsc}.
The configuration data is sent in UDP frames via Ethernet cable from the PC to the FEC.
The FEC can determine the destination of the configuration based on the different UDP ports of the configuration frame.
For the data acquisition, two solutions are available.
The \emph{ESS DAQ}~\cite{essdaq} provides event reconstruction and online monitoring but has limited rate-capabilities when the full non-reconstructed hit data is written to disk (so-called debug mode).
Alternatively, the \emph{tcpdump}~\cite{tcpdump} utility can directly read the data at the network adapter.
The network packages are then written to disk, but online monitoring and visualisation are not possible.
Thus, the solution for high-rate data-taking is to use \emph{tcpdump} for the data acquisition,
and the \emph{ESS DAQ} only for the online monitoring.
This combination allows to obtain the highest possible rates with VMM3a and SRS.
It should be noted that when using standard PC operating systems for high
speed UDP network communication several techniques need to be applied to
minimise the processing overhead and reduce avoidable packet loss \cite{morten}
The offline data analysis \emph{vmm-sdat}~\cite{vmm-sdat} is able to cope with the data from \emph{tcpdump} and \emph{ESS DAQ}.
It clusters the hit data, time matches the clusters in between the readout planes and
reconstructs thus the interaction point of the particle in the gas volume of the detector.
This way the rate of the detected particles can be calculated.
The bit rates on the other hand were directly measured on the network adapter with the Linux command-line tool \emph{ifstat}.
With the DAQ PC\footnote{For the measurements (e.g. the ones presented in section \ref{sec:x-ray-tests}),
a typical 2.5-inch SATA ($\SI{6}{Gbps}$) SSD combined with a $\SI{10}{Gbps}$ network card was used.
No specific optimisation/modification of the software or hardware components was applied.}
as last stage of the readout scheme, the maximum rates per readout stage, which could be achieved,
are summarised in table \ref{tab:results-vs-expectation}.
\begin{table}[t!]
	\centering
	\caption{Rate-capability of the various stages of VMM3a/SRS in the current implementation
	    of the non-ATLAS continuous readout mode.
	    The values given in number of hits/active
	    channels per second ($\si{Mhits/s}$) and data bandwidth in megabits per second ($\si{Mbps}$).
	    The rate at the DAQ computer depends on the number $n$ of network ports on the
	    switch and the bandwidth limit of the network card (e.g. $\SI{10}{Gbps}$).}
	\begin{tabular}{cccc}
		\toprule
		\multicolumn{2}{c}{Readout stage} &
		\multicolumn{2}{c}{Maximum rate}\\
		\multicolumn{2}{c}{\textit{Unit Quantity}} & ($\si{Mhits/s}$) & ($\si{Mbps}$)\\
		\midrule
		\midrule
		\multirow{2}{*}{\textbf{(A)}} & VMM3a channel & 
		\multirow{2}{*}{\num{3.6}} & \multirow{2}{*}{\num{144}}\\
		& \textit{Rate per channel} &   &  \\
		\midrule
		\multirow{2}{*}{\textbf{(B)}} & VMM3a to Spartan-6 &
		\multirow{2}{*}{\num[parse-numbers = false]{8.\overline{8}}} & \multirow{2}{*}{\num[parse-numbers = false]{355.\overline{5}}}\\
		& \textit{Rate per VMM3a} &   &  \\
		\midrule
		\multirow{2}{*}{\textbf{(C)}} & Spartan-6 &
		\multirow{2}{*}{\num[parse-numbers = false]{17.\overline{7}}} & \multirow{2}{*}{\num[parse-numbers = false]{711.\overline{1}}}\\
		& \textit{Rate per HDMI-SerDes} &   & \\
		\midrule
		\multirow{2}{*}{\textbf{(D)}} & HDMI & 
		\multirow{2}{*}{\num[parse-numbers = false]{17.\overline{7}}} & \multirow{2}{*}{\num[parse-numbers = false]{888.\overline{8}}}\\
		& \textit{Rate per Hybrid} &   & \\
		\midrule
		\multirow{2}{*}{\textbf{(E)}} & DVMM and FEC & 
		\multirow{2}{*}{\num[parse-numbers = false]{71.\overline{1}}} & \multirow{2}{*}{\num[parse-numbers = false]{7111.\overline{1}}}\\
		& \textit{Rate per DVMM card} &   & \\
		\midrule
		\multirow{2}{*}{\textbf{(F)}} & Gigabit Ethernet & 
		\multirow{2}{*}{20.8} & \multirow{2}{*}{\num{1000}}\\
		& \textit{Rate per FEC} &   & \\
		\midrule
		\multirow{2}{*}{\textbf{(G)}} & DAQ computer & 
		\multirow{2}{*}{$\num{20.8} \times n$} & \multirow{2}{*}{$\num{1000} \times n$}\\
		& \textit{Rate per Switch Port} &   &  \\
		\bottomrule
	\end{tabular}
	\label{tab:results-vs-expectation}
\end{table}

\section{Measurements with X-rays}
\label{sec:x-ray-tests}
For the rate measurements in the previous descriptions,
test pulses have been used, due to their constant and well-known input rate.
In the following, the results of rate measurements
under application conditions are presented.
First, the experimental set-up is described (section \ref{sec:x-ray-tests_set-up}),
followed by the measurements on the rate limitations (section \ref{sec:x-ray-tests_results-limits}).
As last part, a few examples of high-rate X-ray imaging are presented, demonstrating
the capabilities of VMM3a/SRS (section \ref{sec:x-ray-tests_applications}).

\subsection{Experimental methods}
\label{sec:x-ray-tests_set-up}
For the measurements, a COMPASS-like triple-GEM detector \cite{compass}, which was operated
at a gain of around $\num{e4}$, was uniformly irradiated with X-rays.
The anode featured an $x$-$y$-strip readout with $\num{256}$ strips of $\SI{400}{\micro m}$ pitch
in each direction and an active area of $\num{10} \times \SI{10}{cm^2}$.
As X-ray source, a copper target X-ray tube (Ital Structures Compact 3K5 X-Ray Generator) was used.
The acceleration voltage was set to $\SI{20}{kV}$, while the X-ray tube current was varied.

The electronics set-up was chosen in a way that allowed to achieve the highest readout rates.
For this, only a single hybrid was connected to one FEC, with two FECs in total being
used; one for the readout of the $x$-strips of the detector and one for the readout of the
$y$-strips of the detector.
Hence, the actively read out detector area was only $\num{5} \times \SI{5}{cm^2}$,
with the rest of the area being shielded to absorb the other X-ray photons.

However, most measurements have been performed with the neighbouring-logic of the VMM enabled,
which slightly reduces the rate-capability.
As explained in section \ref{sec:vmm3a-asic}, the neighbouring-logic (NL) allows to read out
the adjacent channels below threshold, if the triggering channel surpassed the threshold.
This feature was enabled because it is beneficial for the position reconstruction and thus
X-ray imaging applications \cite{neighbouring}.


\subsection{Results of the rate limitation measurements}
\label{sec:x-ray-tests_results-limits}
\begin{figure}[t]
	\centering
	\includegraphics[width = 0.9\columnwidth]{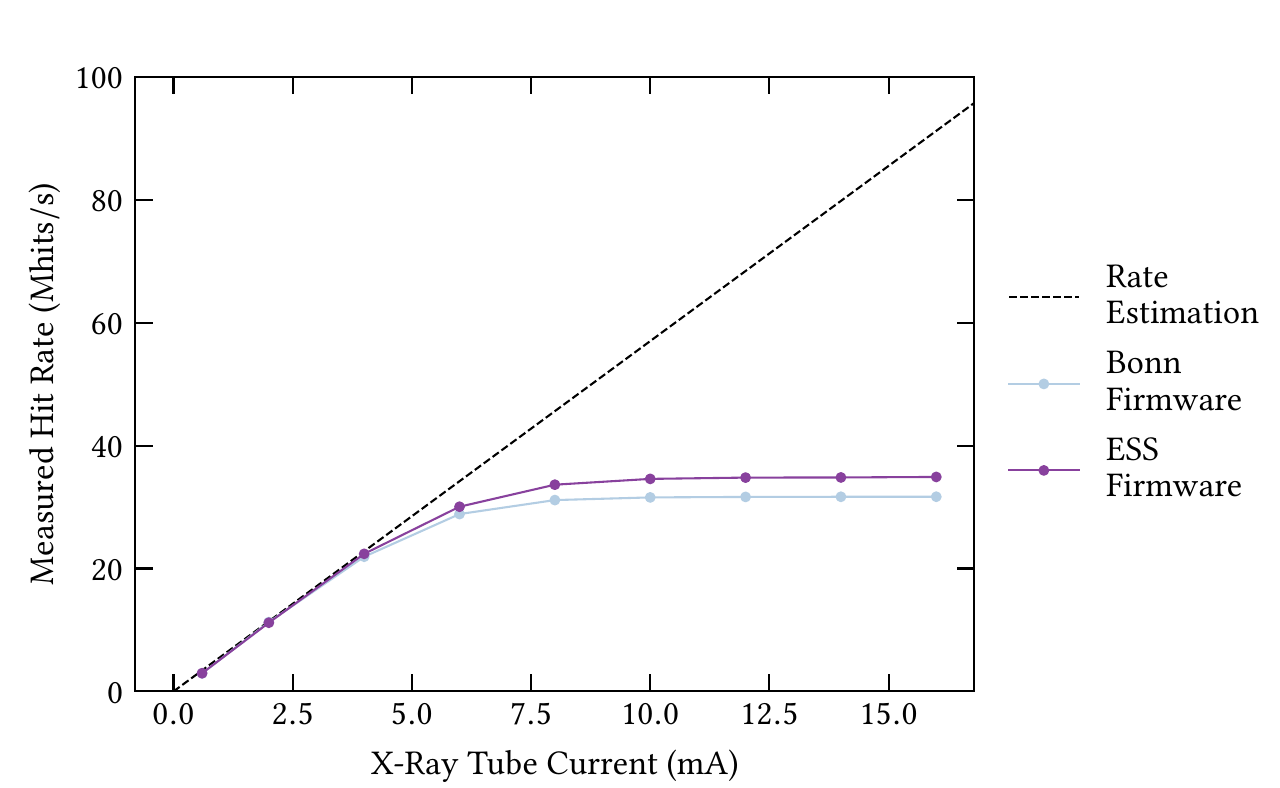} 
	\caption{Measured hit rate (number of active channels in the full set-up per second) depending on
	    the X-ray tube current for the two firmware types.}
	\label{fig:xray_hit-rate}
\end{figure}

\begin{figure}[t!]
	\centering
	\includegraphics[width = 0.9\columnwidth]{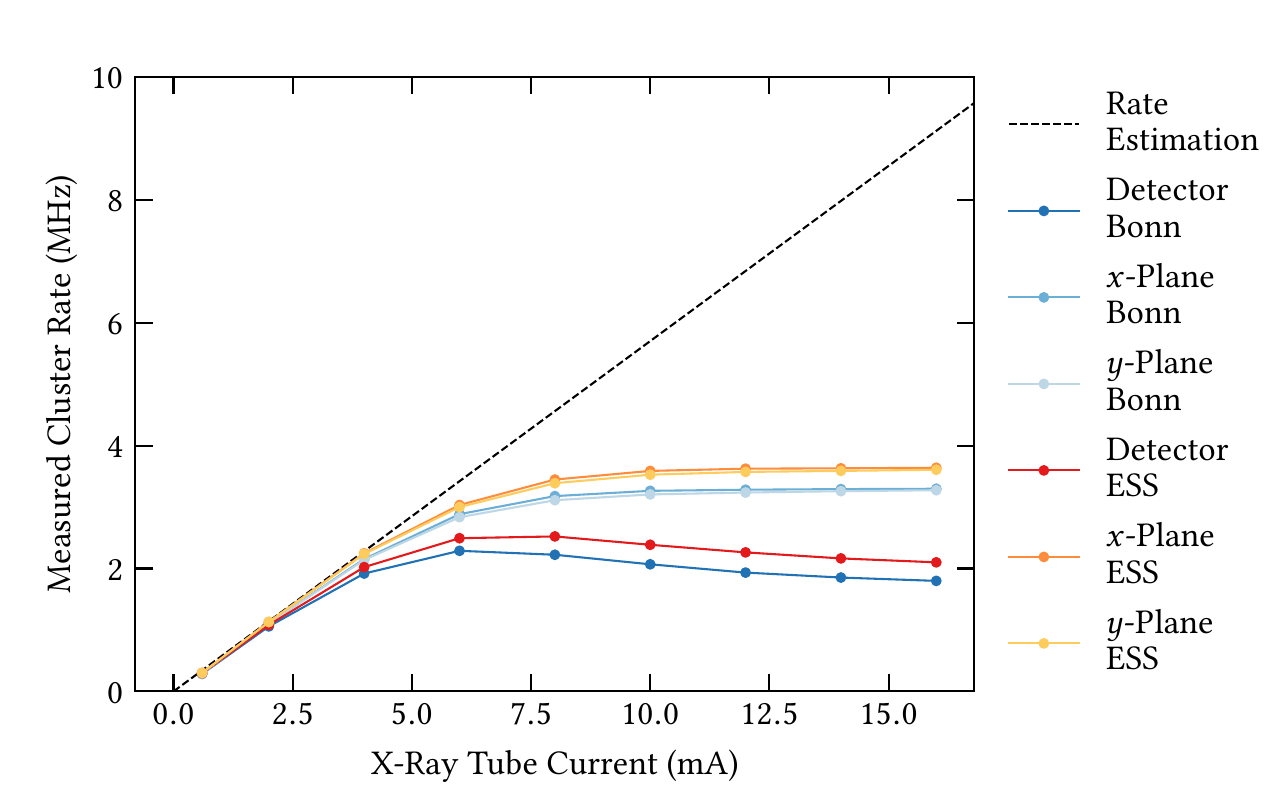} 
	\caption{Measured cluster rate depending on the X-ray tube current for the two firmware types.
	    Both, the cluster rates in each readout plane, as well as the detector cluster rate (where a cluster in the
	    $x$-plane could be matched with a cluster in the $y$-plane), are shown.}
	\label{fig:xray_cluster-rate}
\end{figure}

\begin{figure}[t!]
	\centering
	\includegraphics[width = 0.9\columnwidth]{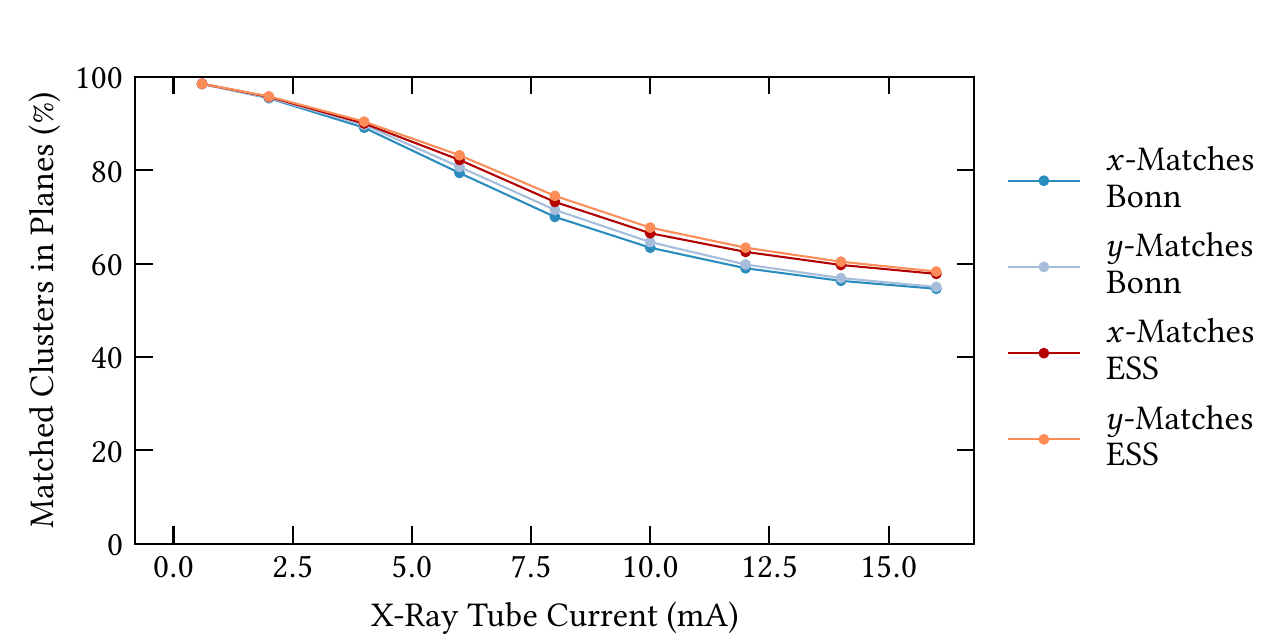} 
	\caption{Fraction of matched clusters in $x$- and $y$-direction
	    depending on the X-ray tube current for the two firmware types.}
	\label{fig:xray_matching-rate}
\end{figure}

The measurement results are shown in figures \ref{fig:xray_hit-rate} to \ref{fig:xray_matching-rate}.
In figure \ref{fig:xray_hit-rate} the measured hit rate, the number
of recorded hits (active channels) per second, is plotted against the X-ray tube current.
A larger X-ray tube current means that more X-ray photons are generated and thus
more hits are expected\footnote{The expected hit and cluster rates are illustrated as dashed
lines in figures \ref{fig:xray_hit-rate} and \ref{fig:xray_cluster-rate}.
The rates are estimated based on the known linear relation between photon flux
and X-ray tube current in the selected current region for the specific X-ray tube.
The linear behaviour is confirmed with a fit for the first three data points
in the low X-ray tube current region, where both firmware versions lead to similar cluster rates
and no cluster losses (plane clusters) are observed.
In addition, for the data point at $\SI{0.6}{mA}$ X-ray tube current, the same rate was
measured with an oscilloscope, while reading the induced signals from the bottom of the last GEM in the
detector.
Based on this, the linear cluster rate behaviour was extrapolated from the first three data points
towards larger X-ray tube currents.
To get the hit rate, the average cluster size of about 5 strips per readout plane
is used (measured at $\SI{0.6}{mA}$).}.
It can be seen that the measured hit rate saturates at high X-ray currents,
indicating a bandwidth limitation.
The cause for the saturation was identified to be the token passing of the VMM3a itself
in combination with the randomly\footnote{Random in position and time,
but still uniformly distributed.} occurring X-ray interactions (see also
section \ref{sec:bonn-hybrid-firmware}).

In figure \ref{fig:xray_cluster-rate} and \ref{fig:xray_matching-rate} the effect of
the saturation of the measured hit rate on the cluster reconstruction is shown;
a cluster contains all hits that belong to the interaction of a single X-ray photon in the detector.
It should be noted that some of the observed effects (see especially figure \ref{fig:xray_matching-rate}) are related
to the $x$-$y$-strip readout\footnote{Here with a 1:1 mapping, meaning that the strip number of the
detector is the same as the readout channel number of the electronics.} and they may be different
for other readout geometries.
Due to the saturation in the hit rate, also a saturation in the cluster rate is expected.
This is observed for the clusters in each individual readout plane (see figure \ref{fig:xray_cluster-rate}).
However, for the detector cluster rate (where a cluster in the $x$-plane could be matched with a cluster in the $y$-plane),
a decrease after reaching a maximum is observed.
The saturation of the measured cluster plane rate can be entirely explained with the bandwidth limits,
while the decrease in the measured cluster detector rate is a combination of the bandwidth limit and the cluster reconstruction.
For its explanation, the cluster reconstruction procedure is recalled.
As mentioned in section \ref{sec:daq-pc}, the clusters are first reconstructed in each readout-plane
individually.
For this, the hits in a certain time interval (here $\Delta t_\mathrm{hits} \leq \SI{150}{ns}$) are considered
as a cluster in time.
Then the geometrical component is added, meaning that the hits in the time cluster occur on adjacent strips
(in the present analysis, one missing strip was allowed).
Hence, a saturation in the hit rate leads to a saturation of the measured cluster rate in the readout plane.
Afterwards, the timestamps of the clusters found in each plane (calculated via centre-of-gravity from the individual
hits, with the charge as weight) are matched in time again (here $\Delta t_\mathrm{plane} \leq \SI{150}{ns}$).
If now, due to bandwidth limitations, data (meaning clusters) are lost/not processed; for example the hits of a cluster
in the $x$-plane can still be reconstructed to a cluster, while the corresponding hits in the $y$-plane are lost due to the
bandwidth limits and thus no combined detector cluster can be identified.
This matching procedure gets less efficient, the larger the actual X-ray interaction rate in the detector gets,
because it means that more hits can be lost.
This explains the decrease in the rate of reconstructed clusters and is illustrated in figure \ref{fig:xray_matching-rate},
where the fraction of the matched clusters for each plane is shown.

\begin{figure}[t!]
	\centering
	\includegraphics[width = 0.9\columnwidth]{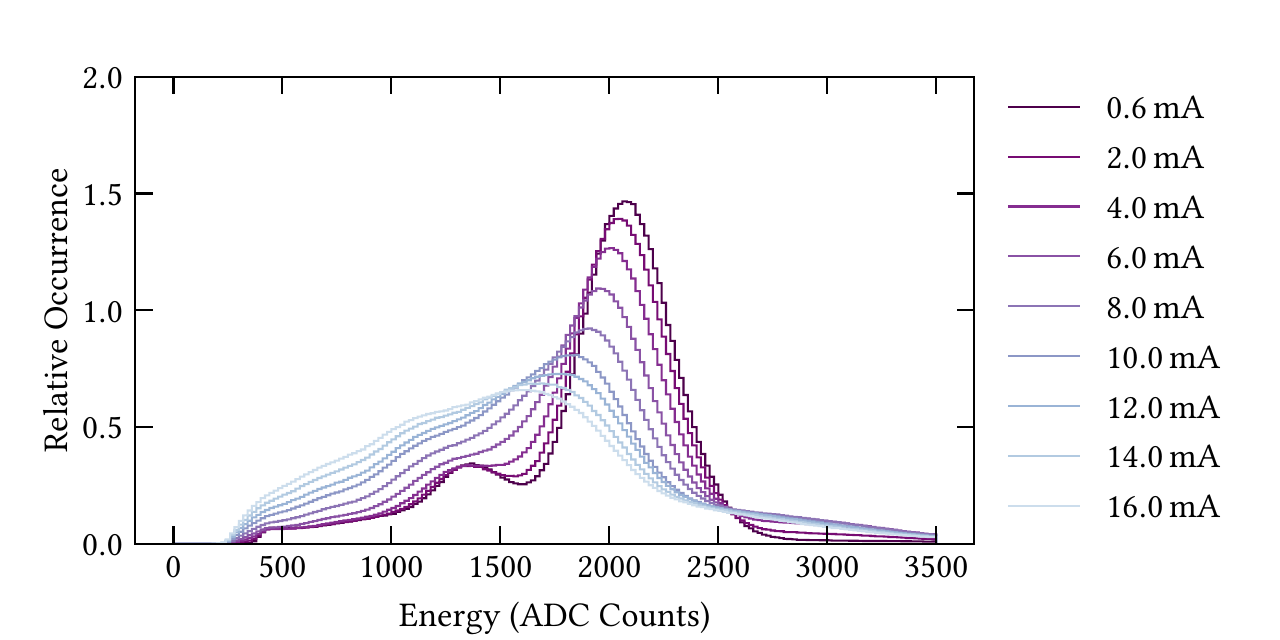} 
	\caption{Measured spectra from the copper target X-ray tube for various X-ray tube currents.}
	\label{fig:xray_spectrum}
\end{figure}

\begin{figure}[t!]
	\centering
	\includegraphics[width = 0.65\columnwidth]{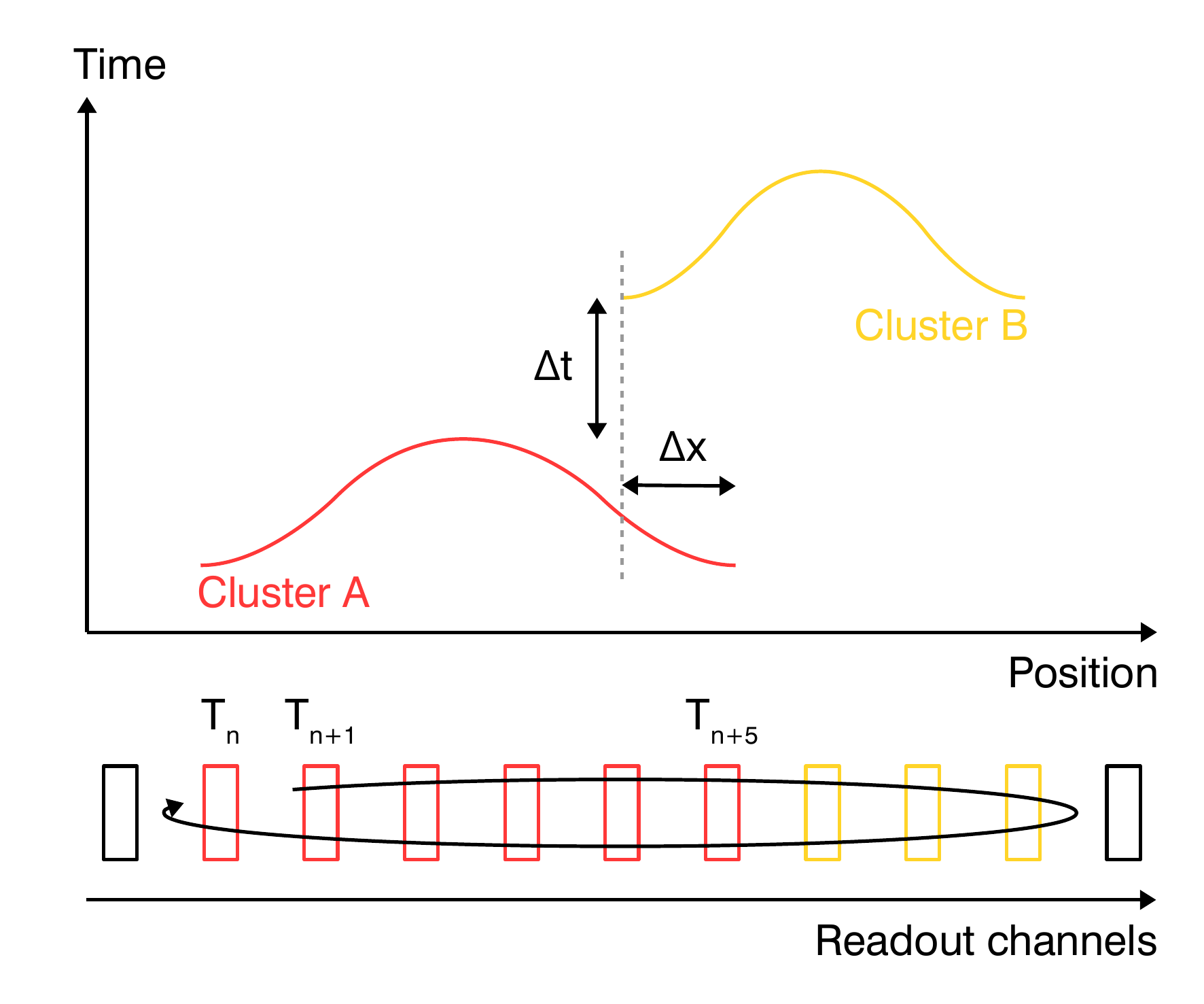} 
	\caption{Illustration of two clusters $A$ and $B$ that have a geometrical
	    overlap of $\Delta x$.
	    The time difference between the occurrence of the clusters is $\Delta t$.
	    The active readout channels are read out token-by-token, with the readout time $T_n$
	    for the token on channel $n$.}
	\label{fig:xray_token-looping}
\end{figure}

In addition to the quantitative effects (fewer reconstructed clusters due
to the reduced cluster finding efficiency), the bandwidth limitations also affect
the quality of the reconstructed data. Figure \ref{fig:xray_spectrum} displays the X-ray spectra for each current setting of the X-ray tube.
The quality of the spectra decreases significantly with increasing hit rates
and is particularly bad, as soon as the bandwidth saturation ($\geq\SI{8}{mA}$ X-ray current)
is reached. The bandwidth limitation originates from the readout of the VMM3a
on the hybrid level\footnote{A bandwidth problem of the FEC can be excluded.
	A hit rate of maximally $\SI{17.5}{Mhits/s}$ per FEC is well below the maximum of
	$\SI{20.8}{Mhits/s}$ derived from the gigabit Ethernet.}.
The readout of each channel is achieved by `looping' over all $\num{64}$ VMM channels
and reading one channel after another that has the active token flag for
stored data; as mentioned in section \ref{sec:vmm3a-hybrid},
a VMM channel can only register a new hit after the previous hit has been read out.
This can lead to the loss of charge information required to reconstruct
the full cluster, as illustrated in figure \ref{fig:xray_token-looping}.
To illustrate the effect of this VMM3a readout limitation, two charge clouds/clusters are shown. The two clusters overlap geometrically and are close in time. Each cluster will generate hits on several readout channels.
The rate limitation occurs, if the time difference $\Delta t = t_B - t_A$
between the occurrence of two clusters $A$ and $B$ is now smaller than
the time difference between the readout of the channels
$\Delta t' = T_{n+1} - T_n$. Some of the hits in the first cluster $A$ (in figure \ref{fig:xray_token-looping} it is channels $n+4$ and $n+5$) have not been read out when the second cluster $B$ occurs. In this case, the information that is acquired by the VMM channels $n+4$ and $n+5$ cannot be stored in the channel buffer and is simply lost.
The more channels are activated, the more likely this type of data loss is to occur due to the token passing scheme and the serial nature of the data transfer.

\begin{figure}[t!]
	\centering
	\includegraphics[width = 0.969\columnwidth]{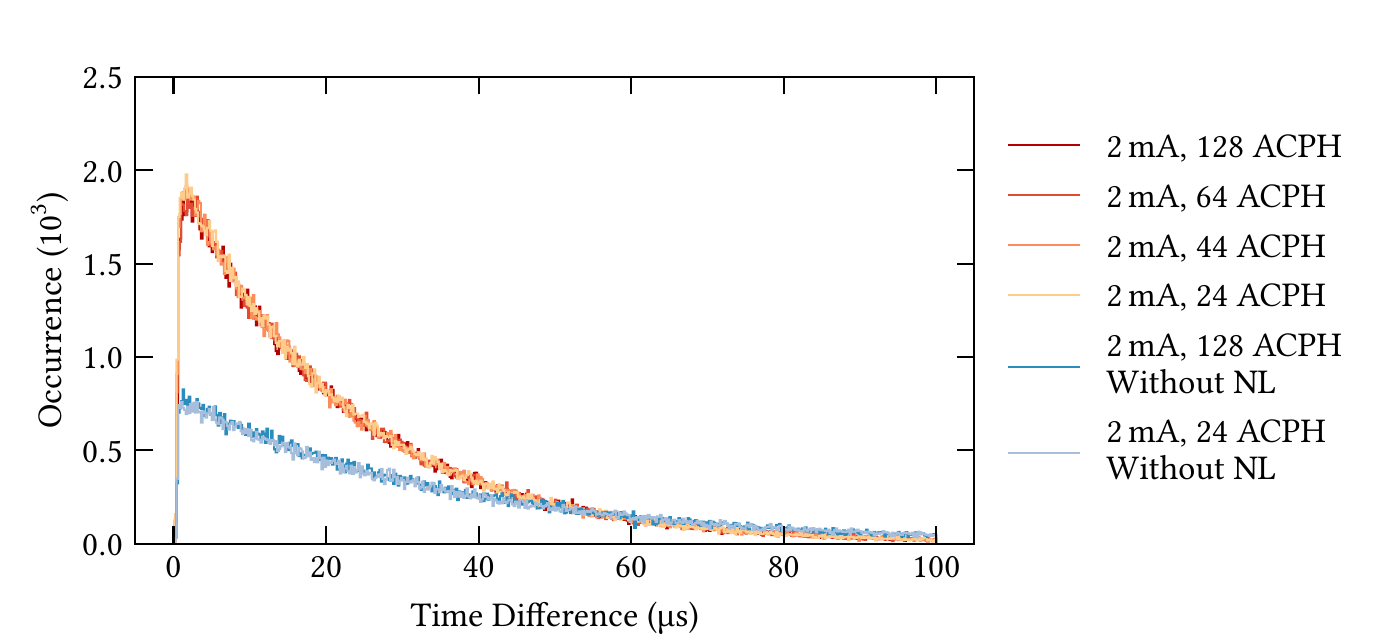}\\ 
	{\footnotesize \textbf{(a)} $\SI{2}{mA}$ X-ray tube current}\\
	\includegraphics[width = 0.969\columnwidth]{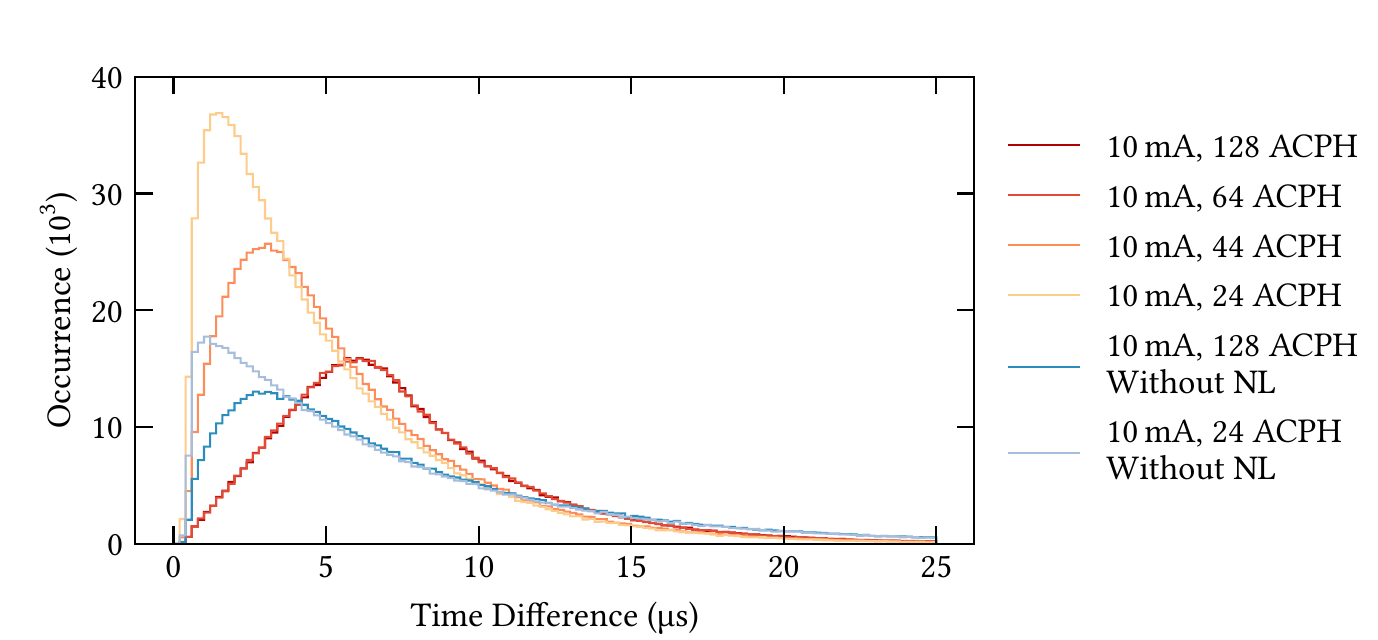}\\ 
	{\footnotesize \textbf{(b)} $\SI{10}{mA}$ X-ray tube current}\\
	\caption{Distributions of the time differences between two consecutive hits for a certain number
	    of Activated Channels Per Hybrid (ACPH) and different X-ray tube currents.
	    The different axis scales between (a) and (b) should be noted.}
	\label{fig:xray_timedifferences}
\end{figure}

This explanation is confirmed by the distributions, shown in figure \ref{fig:xray_timedifferences}.
There, the distributions of the time difference between two consecutive hits on a single channel
are shown, for two X-ray tube currents (meaning different interaction rates) and for various numbers
of Activated Channels Per Hybrid (ACPH).
The time it takes to read out one channel and jump to the next one with the raised token flag
is constant.
Thus, the more channels are active, the more time it takes to loop over all channels with an active token
and read them out.
For a low X-ray current ($\SI{2}{mA}$), the shape of the distributions remains the same, independent
of the number of channels that could be read out.
By turning the NL off, only the number of hits changes as expected
(less active strips per cluster), but not the overall behaviour.
At higher X-ray tube currents ($\SI{10}{mA}$), in the bandwidth limitation,
the effect of the token passing can be seen.
The lower the number of channels which can be read out (ACPH), the quicker one `readout loop' of the
token is finished and the more hits can be read out in the same amount of time.
This can be seen, as the peak of the time difference distributions moves towards
smaller values for smaller ACPHs,
but also for the same number of ACPH, when the NL is turned off.




\subsection{X-ray imaging examples}
\label{sec:x-ray-tests_applications}
As last part of the X-ray studies, a few measurements highlighting the high-rate
imaging capabilities of VMM3a/SRS are shown.
The X-ray tube settings were slightly changed compared to the previous measurements:
the acceleration voltage was set to $\SI{18}{kV}$, while the tube current was kept constant
at $\SI{5}{mA}$.

\begin{figure}[t!]
	\centering
	\includegraphics[width = 0.762\columnwidth]{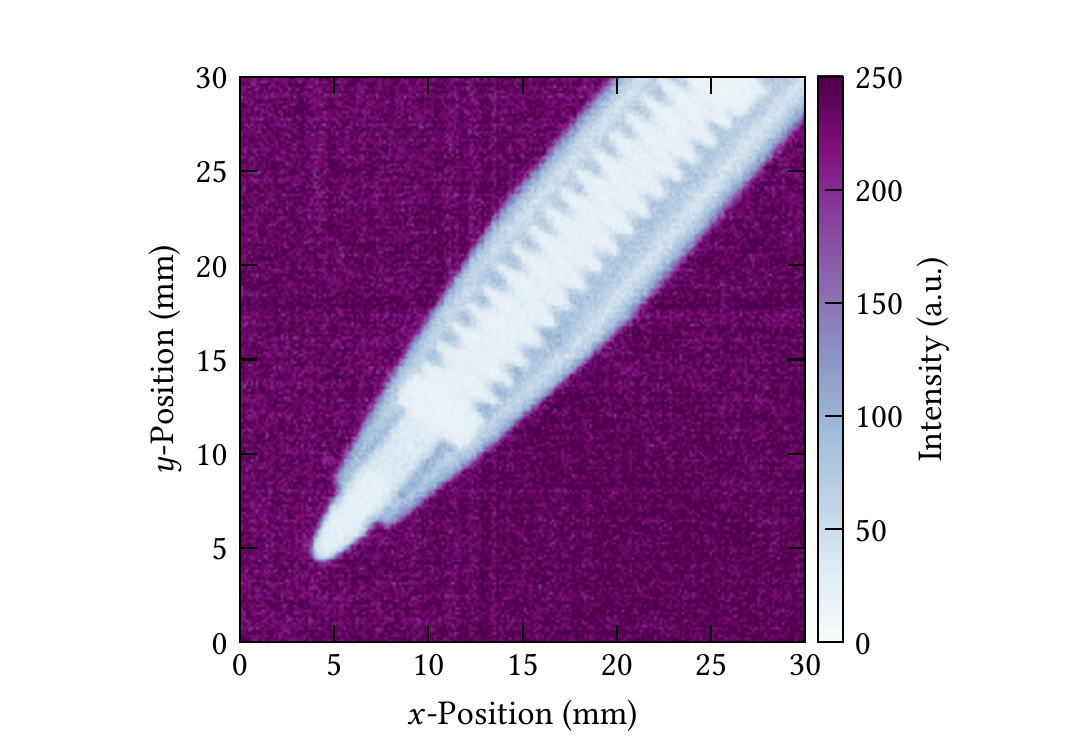} 
	\caption{Image of a pen containing $\num{17e6}$ clusters. The full data set contains $\num{50e6}$ clusters, that have been recorded in
		$\SI{30}{seconds}$.}
	\label{fig:imaging-pen}
\end{figure}

\begin{figure}[t!]
	\centering
	\includegraphics[width = 0.762\columnwidth]{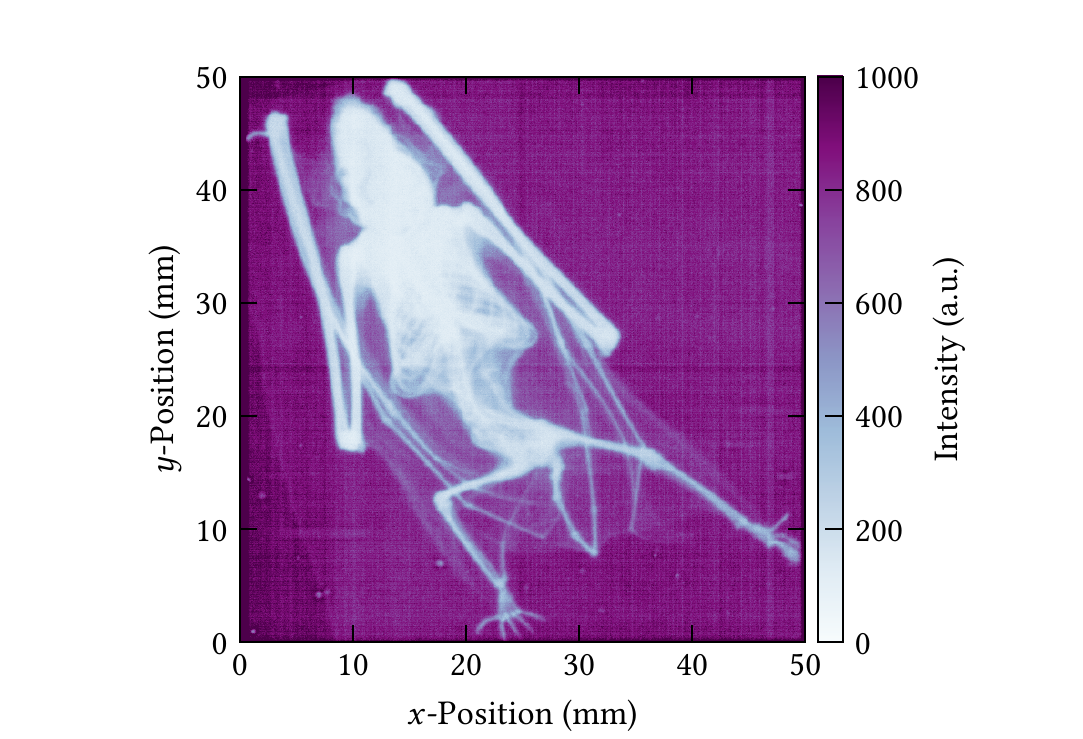} 
	\caption{Image of a dead mammal. The data set for this image
	    contains $\num{277e6}$ clusters, that have been recorded in
		$\num{180}$ seconds.}
	\label{fig:imaging-bat}
\end{figure}

In figure \ref{fig:imaging-pen} an X-ray image of a pen is shown, containing
$\num{17e6}$ clusters.
The full data set, due to the larger active area,
consists of $\num{50e6}$ reconstructed clusters,
which have been recorded in $\num{30}$ seconds, corresponding to a measured
interaction rate of $\SI{1.7}{MHz}$.
The measured hit rate was on average $\SI{20.8}{Mhits/s}$.
Similar to the image of the pen, an X-ray image of a small mammal
was taken (figure \ref{fig:imaging-bat}).
This image was recorded with similar hit and measured cluster rates
as the one of the pen.
However, the acquisition time was slightly longer with $\SI{3}{minutes}$, allowing
to record $\num{277e6}$ clusters for the image.

\begin{figure}[t!]
	\centering
	\includegraphics[width = 0.762\columnwidth]{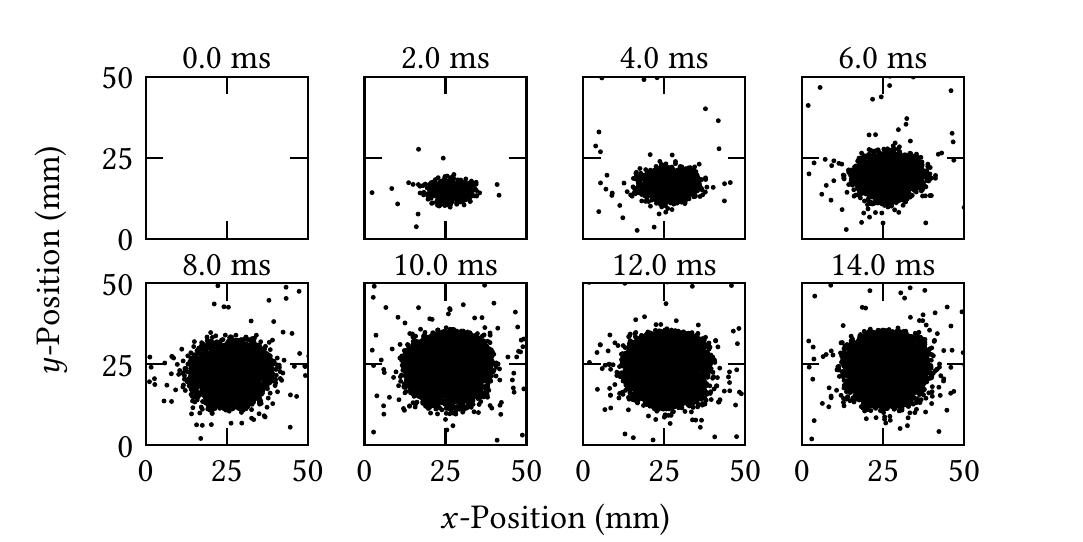}\\ 
	{\footnotesize \textbf{(a)} Frames of the recorded opening of the X-ray tube's shutter.}\\
	\includegraphics[width = 0.762\columnwidth]{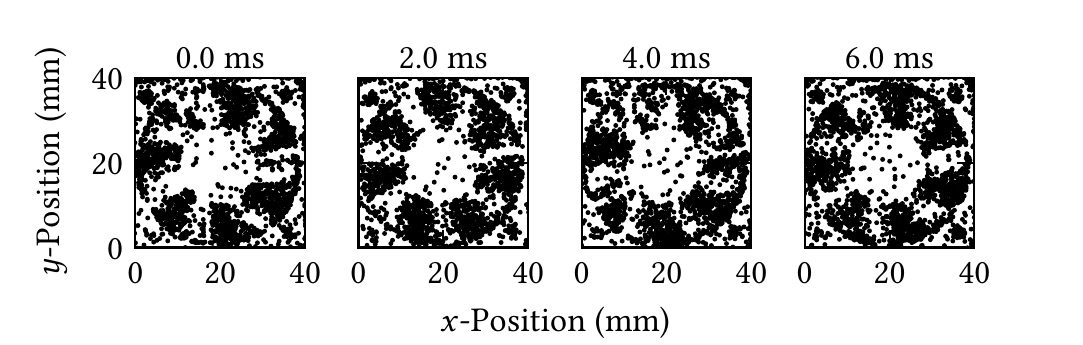}\\ 
	{\footnotesize \textbf{(b)} Frames of the rotating blades of a fan.}\\
	\caption{Examples of the continuous data stream, which is sliced into frames,
	    allowing to reconstruct dynamic processes.}
	\label{fig:frames}
\end{figure}

Due to the high rate-capability with the self-triggered readout and the
individual event reconstruction, it is also possible
to perform imaging of dynamic processes.
To illustrate this, two examples are shown in figure \ref{fig:frames}.
Figure \ref{fig:frames}a displays the opening of the X-ray tube's shutter.
The length of each frame was here chosen to be $\SI{2}{ms}$.
In figure \ref{fig:frames}b, the rotating blades of a fan are shown, again with
a frame length of $\SI{2}{ms}$.
Each frame of the fan contains around $\num{4000}$ clusters.
To increase the absorption of the X-ray photons, the blades have been covered with copper tape. Further, the X-ray tube voltage was reduced to $\SI{16}{kV}$ in order to decrease the Bremsstrahlung fraction in the spectrum.

\section{Conclusion and outlook}
In the past years, the VMM3a ASIC, which was specifically designed to read out gaseous
detectors of the ATLAS New Small Wheel, has been successfully integrated into the RD51 Scalable Readout System. Two VMM3a ASICs are combined together with a Spartan-6 FPGA on a front-end board, the RD51 VMM hybrid.
As part of the integration process, the non-ATLAS continuous self-triggered readout scheme between
the VMMs and the FPGA, as well as the readout scheme towards the Front-End Concentrator (FEC) card, were optimised
for high rates.
In this article, two firmware implementations for the Spartan-6 FPGA have been presented.
Additionally, the optimised firmware on the FEC is described.
It was shown, that readout rates of up to $\SI[parse-numbers = false]{8.\overline{8}}{Mhits/s}$ per VMM
and $\SI{20.8}{Mhits/s}$ per FEC can be achieved.
The results were then confirmed by using the electronics for high-rate X-ray-imaging studies, where in an optimised
set-up, interaction rates of around $\SI{2}{MHz}$ were measured. 

The SRS and specifically the RD51 VMM3a are thus suitable for various applications that require high data rates. Future studies like the measurement of fast charge-up effects in detectors in high irradiation environments, or the multi-channel investigation of space charge effects are now possible.
In the coming productions of the RD51 hybrid, the Spartan-6 FPGA on the RD51 VMM hybrid will be replaced with a Spartan-7 FPGA. A working prototype of the Spartan-7 version already exists. This hardware upgrade will also allow the investigation and implementation of other VMM readout modes, which are e.g. described here \cite{vmm-atlas}. Implementations of even faster readout rates of about $\SI{10}{Mhits/s}$ can be envisaged.

\section*{Acknowledgements}
Parts of this work have been sponsored by the Wolfgang Gentner Programme
of the German Federal Ministry of Education and Research (grant no. 05E18CHA).

Parts of this work have received funding from the European Union’s Horizon 2020
research and innovation programme under the Marie Sklodowska-Curie grant agreement No. 846674
as well as from the BMBF (Germany) under 05K19PD1.


\bibliography{references}

\end{document}